\documentclass{article}

\usepackage{graphicx}
\usepackage{graphics}
\usepackage{amsmath,amssymb,amsfonts}
\usepackage{epsfig}
\usepackage{wrapfig}
\usepackage{amsbsy}
\usepackage{caption}
\usepackage{subcaption}

\usepackage{anysize}
\papersize{29.7cm}{21cm} 
\usepackage{sistyle}
\usepackage{pdfsync}
\usepackage{hyperref}

\newcommand{\bRp}{\mathbb{R}_+}
\newcommand{\bR}{\mathbb{R}^3}
\newcommand{\bx}{\mathbf{x}}
\newcommand{\bv}{\mathbf{v}}
\newcommand{\bc}{\mathbf{c}}
\newcommand{\bu}{\mathbf{u}}
\newcommand{\bg}{\mathbf{g}}
\newcommand{\bG}{\mathbf{G}}
\newcommand{\bom}{\boldsymbol{\omega}}
\newcommand{\cnst}{C}

\newcommand{\md}{\mathrm{d}}
\newcommand{\lp}{\left(}
\newcommand{\rp}{\right)}
\newcommand{\vfi}{\varphi(I)}
\newcommand{\beqs}{\begin{equation*}}
\newcommand{\eeqs}{\end{equation*}}
\newcommand{\beq}{\begin{equation}}
\newcommand{\eeq}{\end{equation}}
\newcommand{\dstar}{ \mathrm{d}r \, \mathrm{d}R \, \mathrm{d}I_* \, \mathrm{d}\mathbf{v}_*}
\newcommand{\dsvi}{ \mathrm{d}r \, \mathrm{d}R \, \mathrm{d}I_* \, \mathrm{d}\mathbf{v}_* \, \mathrm{d}I \,  \mathrm{d}\mathbf{v}}
\newcommand{\dsvic}{ \mathrm{d}r\, \mathrm{d}R\, \mathrm{d} I_* \, \mathrm{d}\mathbf{c}_* \, \mathrm{d}I \,  \mathrm{d}\mathbf{c}}
\newcommand{\dsvig}{ \mathrm{d}r \, \mathrm{d}R \, \mathrm{d}I_* \, \mathrm{d}\mathbf{G} \, \mathrm{d}I \,  \mathrm{d}\mathbf{g}}
\newcommand{\inzv}{\left|\,\mathbf{v} \right|}
\newcommand{\inzvs}{\left| \mathbf{v}_* \right|}
\newcommand{\inzvp}{\left| \mathbf{v}' \right|}
\newcommand{\inzvps}{\left| \mathbf{v}'_* \right|}
\newcommand{\inzvMvs}{\left| \mathbf{v} - \mathbf{v}_* \right|}
\newcommand{\inzvpMvps}{\left| \mathbf{v}' - \mathbf{v}'_* \right|}
\newcommand{\koren}{\sqrt{\frac{R \, E}{m}}}
\newcommand{\Tom}{T_{\boldsymbol{\omega}} \left[  \frac{\mathbf{v} - \mathbf{v}_*}{\left| \mathbf{v} - \mathbf{v}_* \right|} \right] }
\newcommand{\Tomc}{T_{\boldsymbol{\omega}} \left[ \frac{\mathbf{c} - \mathbf{c}_* }{\left| \mathbf{c} - \mathbf{c}_* \right|} \right]}
\newcommand{\z}{\zeta_0(T)}

\newtheorem{lemma}{Lemma}
\newtheorem{proposition}{Proposition}

\begin{document}

\title{Polyatomic gases with dynamic pressure:\\ Maximum entropy principle and shock structure}

\author{Milana Pavi\'{c}-\v{C}oli\'{c}, Damir Madjarevi\'{c}, Srboljub Simi\'{c}}

\maketitle

\begin{abstract}
This paper is concerned with the analysis of polyatomic gases within the framework of kinetic theory. Internal degrees of freedom are modeled using a single continuous variable corresponding to the molecular internal energy. Non-equilibrium velocity distribution function, compatible with macroscopic field variables, is constructed using the maximum entropy principle. A proper collision cross section is constructed which obeys the micro-reversibility requirement. The source term and entropy production rate are determined in the form which generalizes the results obtained within the framework of extended thermodynamics. They can be adapted to appropriate physical situations due to the presence of parameters. They are also compared with the results obtained using BGK approximation. For the proposed model the shock structure problem is thoroughly analyzed.
\end{abstract}

\section{Introduction}

Non-equilibrium processes in polyatomic gases are peculiar since there appears dynamic pressure as an excess normal pressure added to standard thermodynamic pressure. Moreover, experiments showed that the influence of dynamic pressure on transport processes prevails the influence of shear stresses and heat flux. Therefore, a proper description of behaviour of the polyatomic gases, and dynamic pressure, is needed in continuum and kinetic theories alike.

Classical continuum theory used the model of Newtonian fluid which related the dynamic pressure to bulk viscosity and compressibility of the medium. As it is well known, such constitutive relations lead to parabolic governing equations \cite{de2013non}. Although extensively used in practical situations, they predict paradoxical infinite speed of propagation of disturbances. Extended thermodynamics (ET)---a phenomenological theory that fills the gap between macro and meso scale---resolved this paradox by proposing a set basic principles and systematic procedures that led to governing equations in the form of hyperbolic systems of balance laws \cite{RET}. In its basic form ET can be related to the moment method of kinetic theory of monatomic gases \cite{Grad}, although the closure procedure is entirely different. Third approach---application of the maximum entropy principle (MEP) \cite{Kogan}---yields the models that are equivalent to ET and moment method \cite{Dreyer}. It is basically of kinetic character, but fits perfectly into the framework of the so-called molecular extended thermodynamics \cite{RET1}. All three approaches mentioned above were successfully applied to monatomic gases and their mixtures, but failed to provide an extension that naturally incorporates the dynamic pressure into their framework (there were certain attempts, but they were not widely accepted).

Recently, a substantial progress was made in the study of polyatomic gases within ET. A new point of view was proposed, which regards the governing equations as members of two hierarchies, momentum and energy one, in contrast to a single hierarchy typical for monatomic gases \cite{ETdense}. In such a way, trace of the stress (pressure) tensor is no longer related to the internal energy density, and dynamic pressure naturally ``fills'' this free space. The fundamental idea of two hierarchies found its natural counterpart in the MEP for rarefied polyatomic gases, where internal degrees of freedom of the molecules were modelled using single continuous real variable \cite{Pavic}. In other words, molecular internal degrees of freedom implied the existence of dynamic pressure. Subsequent studies of polyatomic gases within ET were summarized in the monograph \cite{RET_Poly}. It also contains the results of application of MEP in this context. Extensive study of molecular extended thermodynamics for polyatomic gases was given in \cite{AMR_MolecularET}.

The case of a polyatomic gas with dynamic pressure, in which other non-equilibrium effects (shear stresses and heat conduction) were neglected, is of particular interest for the analysis. Namely, it was found that it is one of the rare systems that admits a non-linear closure of the governing equations using the entropy principle \cite{ARST_NLET6,TARS_Overshoot,ARST_RecentI}. On the other hand, it admits exact solution to the variational problem of MEP, with convergent moments of the distribution function \cite{Ruggeri_MEP6}. However, in the case of polyatomic gases with 6 fields, application of MEP was limited to the simplest possible case of BGK model of interaction, leading to linear source term in the governing equation for dynamic pressure.

Aim of this paper is to analyze the application of MEP to polyatomic gases with 6 fields. Dynamic pressure will not be introduced \emph{a priori}, but as a measure of deviation of the system from an equilibrium state. The governing equations will be derived using a complete nonlinear collision operator, and appropriate entropy production will be calculated. There will appear the main difference between our approach and the results of non-linear ET---our model is more flexible since it contains more free parameters in the source term, and their values may be adapted to fit into particular physical situation. Similar results were obtained in the case of 14 moments equations for polyatomic gases \cite{Pavic_Poly}, where the presence of parameters admitted the matching with Prandtl number and temperature dependence of viscosity. As an illustration, the problem of shock structure will be analyzed and dependence of the shock profile upon the values of parameters will be discussed. Finally, this study will raise a question of relation between mesoscopic and macroscopic description. Namely, it will be shown that proper choice of the collision cross section---the one that assumes its micro-reversibility and Galilean invariance---yields the macroscopic source term which is not equivalent to the one of non-linear ET. However, if the condition of Galilean invariance is dropped, one may construct the cross section that yields the same source term as non-linear ET. This example cannot provide definite answer, but it points out the phenomenon that the cross section improper at the mesoscopic level, eventually leads to a proper macroscopic description.

The paper is organized as follows. Section 2 contains kinetic description of polyatomic gases with continuous internal energy, including the analysis of collision operator, macroscopic densities, the $H$-theorem and the equilibrium distribution function. In Section 3 an application of MEP to polyatomic gases in local equilibrium will be given, along with macroscopic equations. Section 4 will contain the application of MEP to polyatomic gases with dynamic pressure as a field variable. This will include the derivation of non-equilibrium distribution function, field equations and non-linear source term, analysis of entropy density and entropy production and the introduction of dynamic pressure as a measure of deviation from local equilibrium. At this point, the results of the present study will be compared with the results of ET. In Section 5, an analysis of the shock structure problem will be given. The paper will be closed by concluding remarks and an Appendix that will contain detailed calculation of the source term.

\section{Polyatomic gases with continuous internal energy}

The main feature of a polyatomic gas is the presence of additional degrees of freedom (other than translational ones). This phenomenon can be captured in a variety of ways. We consider the case in which the additional degrees of freedom are modelled by means of  one single continuous (that is, non discrete) parameter $I \in \bRp$, the so-called microscopic internal energy of a molecule, which acts as a lumped parameter for non-translational degrees of freedom \cite{BorgLars,PolyModel}. From the kinetic point of view, the state of a rarefied polyatomic gas is statistically described by a distribution function $f\geq 0$, which depends on the following variables: time $t \in \bRp$, position in the physical space $\bx \in \bR$,  velocity of the particle $\bv \in \bR$, but also on a microscopic internal energy $I \in \bRp$.

The evolution of the distribution function $f := f(t, \bx, \bv, I)$ is determined by the Boltzmann equation:
\beq\label{BE}
\partial_t f + \bv \cdot \nabla_\bx f =  Q(f, f).
\eeq
The right-hand side of the Boltzmann equation, $Q(f,f)$, is a (quadratic) bilinear  operator acting only on the velocity $\bv$ and the internal energy $I$. It depends on the collision model used to describe the interaction between the molecules. Since it will be important for subsequent macroscopic modelling, we analyze it in certain details.

\subsection{Collision transformation}\label{section: coll transf}It is assumed that molecules interact via binary elastic collisions, and thus we  analyze a collision process between two molecules. Suppose that colliding molecules have velocities $\bv'$ and $\bv'_*$ and internal energies $I'$ and $I'_*$, respectively. After the collision they are transformed into post-collisional quantities; velocities become $\bv$ and $\bv_*$, while internal energies transform to $I$ and $I_*$. The conservation laws of momentum and total energy (kinetic plus internal that captures phenomena related to polyatomic gases) of a system consisting of two molecules are valid during a collision process:
\beq\label{microscopic CL}
\begin{split}
m \bv' + m \bv'_* &= m \bv + m \bv_*,\\
\tfrac{m}{2} \inzvp^2 + \tfrac{ m}{2} \inzvps^2 + I' + I'_* &= \tfrac{m}{2}  \inzv^2 + \tfrac{m}{2}  \inzvs^2 + I + I_*.
\end{split}
\eeq
It can be rewritten in the center of mass reference frame, obtaining an equivalent system
\beq\label{micropcopic CL center of mass}
\begin{split}
m \bv' + m \bv'_* &= m \bv + m \bv_*,\\
E = \tfrac{m}{4}   \inzvpMvps^2  + I' + I'_* &= \tfrac{m}{4} \inzvMvs^2  + I + I_*.
\end{split}
\eeq
This is a system of four equations for eight unknowns---the post-collisional velocities $\bv$, $\bv_*$ and post-collisional internal energies  $I$, $I_*$---and it is natural to expect that the solution of this system will be expressed in terms of four parameters. We use the so-called Borgnakke-Larsen procedure, which is based on the repartition of the total energy $E$. At first, the total energy $E$ is divided into translational kinetic energy and microscopic internal energy. We introduce a parameter $R\in [0,1]$ in order to impose a part $R E$ of the total energy as pre-collisional kinetic energy, and the rest $(1-R)E$ to the microscopic internal energy (of the two colliding molecules):
\beq\label{R E}
R E     = \tfrac{m}{4} \inzvpMvps^2, \qquad
(1-R) E = I'+I'_*.
\eeq
Further,  with the help of another parameter $r \in [0,1]$, the internal energy itself is distributed between two molecules:
\beq\label{I', I'_*}
I'=r(1-R)E, \qquad
I'_* = (1-r) (1-R)E.
\eeq
In order to determine the pre-collisional velocities, the equation \eqref{R E} is parametrized with the help of  some unit vector $\bom \in S^2$  as follows:
\beq\label{v' - v'_* parametrizovano sa omega}
\bv'-\bv'_* = 2 \, \koren \, \Tom,
\eeq
where $T_{\bom} \left[  \mathbf{y} \right] = \mathbf{y} - 2 \lp \bom \cdot \mathbf{y} \rp \bom$, for $\mathbf{y} \in \bR$. Combining \eqref{v' - v'_* parametrizovano sa omega} and \eqref{micropcopic CL center of mass}$_1$, we obtain expressions for the  pre-collisional velocities in terms of the post-collisional quantities:
\begin{equation}\label{collisional rules}
\bv'= \frac{\bv+\bv_*}{2} + \koren \, \Tom,\qquad
\bv'_* = \frac{\bv+\bv_*}{2} - \koren \, \Tom.
\end{equation}

For a fixed $\bom \in S^{2}$, we define the collision transformation
\begin{align*}
S_{\bom}: \mathbb{R}^{6}  \times \mathbb{R}^2_+\times [0,1]^2 &\rightarrow \mathbb{R}^{6}  \times \mathbb{R}^2_+\times [0,1]^2\\
(\bv,\bv_*,I,I_*,r,R)&\mapsto(\bv',\bv'_*,I',I'_*,r',R')
\end{align*}
by the relations \eqref{collisional rules}, \eqref{I', I'_*} and
\beq\label{R', r'}
R' = \frac{m}{4\, E} \inzvMvs^2,\qquad
r' = \frac{I}{I + I_*}.
\eeq
We collect its main properties in Lemma \ref{Lemma J non prime -> prime}.
\begin{lemma}\label{Lemma J non prime -> prime}
For any $\bom \in S^{2}$, the transformation $S_{\bom}$ is an involution of the set $\mathbb{R}^{6}  \times \mathbb{R}^2_+\times [0,1]^2$ and its Jacobian determinant is given by
\beq\label{jacobian non prime -> prime}
J_{S_{\bom}} = \frac{(1-R)}{(1-R')}\lp\frac{R}{R'}\rp^{\frac{1}{2}} = \frac{(1-R)}{(1-R')} \frac{\inzvpMvps}{\inzvMvs}.
\eeq
\end{lemma}

\noindent What about the proof of the Lemma?

\subsection{Collision operator}
We consider the following model for the collision operator
\beq\label{PolyA: coll integral}
Q(f,f)(\bv, I) =  \int_{\bR \times \bRp \times [0,1]^2 \times S^{2}} \left[ f' f'_* - f f_* \right]  \mathcal{B} \lp 1-R \rp R^{\frac{1}{2}} \frac{1}{\vfi} \, \md \bom \, \dstar,
\eeq
where the standard abbreviations are used $f':=f(t, \bx, \bv', I')$, $f'_*:=f(t, \bx, \bv'_*, I'_*)$, $f_*:=f(t, \bx, \bv_*, I_*)$, and  $\bv', \bv'_*, I', I'_*$ are defined in \eqref{collisional rules} and \eqref{I', I'_*}. The cross section $\mathcal{B}:=\mathcal{B}(\bv, \bv_*, I, I_*, r, R, \bom)$ is a nonnegative function that satisfies the microreversibility assumptions (reference needed):
\beqs
\mathcal{B}(\bv, \bv_*, I, I_*, r, R, \bom) = \mathcal{B}(\bv', \bv'_*, I', I'_*, r', R', \bom) = \mathcal{B}(\bv_*, \bv, I_*, I, 1-r, R, \bom).
\eeqs

This equation should be numbered.

\begin{proposition}\label{Proposition weak form}
Let $\psi: \bR \times \bRp \rightarrow \mathbb{R}$ be a function of the velocity $\bv$ and microscopic internal energy $I$, such that the integral
\beqs
\int_{\mathbb{R}^3 \times \mathbb{R}_+}  Q(f, f)(\bv, I) \, \psi(\bv, I) \, \varphi(I) \, \md \bv\, \md I
\eeqs
makes sense. Then, the following holds
\begin{multline}\label{polyA: weak form}
\int_{\mathbb{R}^3 \times \mathbb{R}_+} Q(f, f)(\bv, I) \, \psi(\bv, I) \, \varphi(I) \, \md \bv\, \md I  \\= - \frac{1}{4} \int_{ \mathbb{R}^3 \times \mathbb{R}_+ \times \mathbb{R}^3 \times \mathbb{R}_+ \times [0,1]^2 \times S^{N-1}} \left[ f' f'_* - f  f_* \right] \times \left[ \psi(\bv', I') + \psi(\bv'_*, I'_*)- \psi(\bv, I) - \psi(\bv_*, I_*) \right] \\ \times\mathcal{B} \,  \lp 1-R \rp R^{\frac{1}{2}} \,\mathrm{d} \bom \, \dsvi.
\end{multline}
\end{proposition}

Since Proposition 1 will not be proved here, a reference for a proof is needed.

As a consequence of the Proposition, there exist test-functions $\psi(\bv, I)$, that can be chosen independently of $f$, such that (\ref{polyA: weak form}) vanishes. These test-functions constitute a set of collision invariants:
\beq \label{collision invariants}
\lp \begin{matrix}
m \\
m \bv \\
\tfrac{m}{2}\inzv^2 + I
\end{matrix} \rp,
\eeq
(molecular mass $m$ is inessential, but convenient, for the definition of collision invariants) which imply the conservation properties of the weak form of the collision integral:
\beq\label{anuliranje weak form}
\int_{\mathbb{R}^3 \times \mathbb{R}_+} Q(f,f)(\bv, I) \,\lp \begin{matrix}
m \\
m \bv\\
\tfrac{m}{2}\inzv^2 + I
\end{matrix} \rp \vfi \,  \md I \, \md \bv=0.
\eeq
\subsection{$H-$theorem} The entropy production functional in polyatomic gases is defined as:
\begin{equation}\label{entropy production}
D(f):= \int_{\mathbb{R}^3 \times \mathbb{R}_+} Q (f, f)(\bv, I) \, \lp \log  f  \rp  \vfi \, \md I \, \md\bv.
\end{equation}
It is well defined when $f$ is nonnegative and satisfies some suitable assumptions of regularity, lower bound, and decay at infinity (reference needed). The following properties hold.

\begin{proposition}\label{H teorema} Assume that the cross section $\mathcal{B}$ is positive almost everywhere and that $f \geq 0$ is such that both the collisional integral $Q$ and the entropy production $D$ are well defined. Then
\begin{enumerate}
\item[(a)] The entropy production is non-positive, i.e.
$
D(f)  \leq 0.
$
\item[(b)] Moreover, the following three properties are equivalent:
\begin{enumerate}
\item[ i. ] For any $\bv \in \mathbb{R}^3$, $I \in \bRp$,
$
Q (f, f)(\bv, I) = 0;
$
\item[ ii. ] The entropy production vanishes, that is
$
D(f)=  0;
$
\item[ iii. ] There exist $n>0$, $T>0$ and $\bu \in \mathbb{R}^3$ such that
\begin{equation*}
f(\bv, I) = \frac{\rho}{m \z} \left( \dfrac{m}{ 2 \, \pi \, k \, T} \right)^{3/2} e^{-\frac{1}{ k T} \lp \frac{m}{2}|\bv-\bu|^2+ I\rp},
\end{equation*}
where $\z$ is given by:
\beqs
\z = \int_{\bRp}  \varphi(I) \,
    e^{- \frac{1}{k T} I} \, \md I.
\eeqs
\end{enumerate}
\end{enumerate}
\end{proposition}

Proposition 2 represents the $H-$theorem for polyatomic gases (see \cite{PolyModel},\cite{PolyMix} for the proof). Note that the necessary(?) condition for $Q(f,f)(\bv, I)=0$ to hold
is that $f$ is merely local Maxwellian distribution function, i.e.:
\beq\label{local Maxw}
f_E(t, \bx, \bv, I) = \frac{\rho(t, \bx)}{m \zeta_0\lp T (t, \bx) \rp} \left( \dfrac{m}{ 2 \, \pi \, k \, T(t, \bx)} \right)^{3/2} e^{-\frac{1}{ k \,T(t, \bx)} \lp \frac{m}{2}|\bv-\bu(t, \bx)|^2+ I\rp},
\eeq
where the quantities $\rho$, $\bu$ and $T$ are allowed to depend on time $t$ and position $\bx$. This dependence is not arbitrary, but depends on the left-hand side of the Boltzmann equation.

\section{Polyatomic gases in local equilibrium}

The maximum entropy principle is based upon assumption that our information about the system is acquired through macroscopic quantities. Therefore, the most probable velocity distribution function is the one that maximizes the kinetic entropy:
\beq\label{entropy}
h = - k \int_{\bR \times \bRp} f \, \log f \, \vfi \, \md I \md \bv,
\eeq
and is compatible with the chosen set of macroscopic quantities. Having in mind that macroscopic quantities are expressed as moments of the distribution function, it appeared that MEP is an efficient tool for the closure of macroscopic equations derived as transfer equations for the moments. To start with, MEP for polyatomic gases described by 5 fields will be analyzed. The main results have already been given in \cite{Pavic}, but we shall give them a new flavour.

\paragraph{Field variables.} Physically important macroscopic quantities can be interpreted as moments of the distribution function. Integration against collision invariants defines the mass density, the momentum density and the total energy density of a gas:
\beq \label{macroscopic quantites}
\lp \begin{matrix} \rho \\ \rho \, \bu \\  \frac{1}{2}\rho \left| \bu \right|^2 + \rho e \end{matrix} \rp =  \int_{\mathbb{R}^3 \times \mathbb{R}_+} \lp \begin{matrix} m \\ m \, \bv \\    \frac{m}{2} \left| \bv \right|^2 +I   \end{matrix} \rp \, f \, \vfi \, \md I \, \md\bv,
\eeq
where $\bu$ represents the macroscopic velocity and $\rho e$ is the internal energy density of the gas. Introducing the peculiar velocity $\bc$ with respect to the macroscopic velocity $\bu$ by the relation $\bc=\bv-\bu$, definition (\ref{macroscopic quantites}) implies:
\beq \label{intrinsic macroscopic quantites}
\lp \begin{matrix} \rho \\ \mathbf{0} \\  \rho e \end{matrix} \rp =  \int_{\mathbb{R}^3 \times \mathbb{R}_+} \lp \begin{matrix} m \\ m \, \bc \\    \frac{m}{2} \left| \bc \right|^2 +I   \end{matrix} \rp \, f \, \vfi \, \md I \, \md\bv.
\eeq
Mention that,
for the choice of the weight function $\vfi = I^{\alpha}$, the proper classical caloric equation of state is obtained for polyatomic gases in equilibrium, i.e. for $f=f_E$:
\begin{equation} \label{internal energy}
\left. e \right|_{E} = \left(  \alpha + \frac{5}{2}   \right) \frac{k}{m} T; \qquad \alpha > -1.
\end{equation}
Thanks to this identification, parameter $\alpha$ can be related to the heat capacity ratio of the gas $\gamma = \frac{D + 2}{D}$, $D$ being the number of degrees of freedom:
\beqs
\alpha = \frac{-5 \, \gamma + 7}{2 \lp \gamma - 1 \rp}.
\eeqs

\paragraph{Distribution function.} We determine the distribution function corresponding to the equilibrium quantities---mass, momentum and energy density. Therefore, MEP is equivalent to the following  variational problem: determine the distribution function $f:=f(t,\mathbf{x},\bv,I)$ such that
\begin{alignat}{3}\label{MEP eq: var problem}
\text{max}_f& \quad h  \nonumber\\
\text{s.t.}  & \lp \begin{array}{c}
    \rho \\
    0_i \\
    \rho e
  \end{array}\rp   =  \int_{\bR \times \bRp}
     \left(
  \begin{array}{c}
    m \\
   m\, \left| \bc \right|^2\\
   \frac{m}{2} \left| \bc \right|^2 + I
  \end{array}
  \right)\,
    f \, \vfi \, \md I \, \md\bc.
\end{alignat}
\begin{proposition}
The  distribution function which maximizes the entropy \eqref{entropy} subject to constraints \eqref{MEP eq: var problem} for the choice of the weighting function $\varphi(I) = I^{\alpha}$, has the form:
\begin{equation}\label{solution MEP eq}
\hat{f}_5(t, \bx, \bc + \bu, I) = \frac{\rho}{m} \lp \frac{1}{2 \pi e } \left( \alpha + \frac{5}{2} \right)  \rp^{3/2} \lp \frac{1}{m e } \left( \alpha + \frac{5}{2} \right)  \rp^{\alpha+1} \frac{1}{\Gamma\left[ \alpha+1\right]}\,
 e^{-\frac{1}{m e } \left( \alpha + \frac{5}{2} \right) \left( \frac{1}{2} m \left|  \bc \right|^2 +I \right) }
\end{equation}
\end{proposition}

The proof of the Proposition 3 is omitted, since it can be found in \cite{Pavic}. Note that MEP yields the solution that is local, i.e. $\rho$ and $e$ may depend on time $t$ and position $\mathbf{x}$. Furthermore, we retained the original macroscopic quantities (\ref{macroscopic quantites}) in the distribution function and did not introduce the temperature as the state variable, so far. If the internal energy is identified as (\ref{internal energy}), then (\ref{solution MEP eq}) reduces to (\ref{local Maxw}). In other words, MEP for 5 fields yields the local Maxwellian as solution.

\paragraph{Field equations.} Macroscopic variables $\rho$, $\bu$ and $e$ locally depend on $t$ and $\mathbf{x}$. They are not arbitrary; they have to satisfy the macroscopic equations obtained by integration of the Boltzmann equation \eqref{BE} against the weighting functions introduced in \eqref{macroscopic quantites}:
\begin{align} \label{field eqs5}
  & \partial_{t} \rho + \sum_{k=1}^3 \partial_{x_k} (\rho u_{k}) = 0,
\nonumber \\
  & \partial_{t} (\rho u_{i}) + \sum_{k=1}^3 \partial_{x_k}
    (\rho u_{i} u_{k} + p \, \delta_{ik}) = 0,
\\
  & \partial_{t} \left( \tfrac{1}{2} \rho \left| \bu \right|^{2} +
    \rho e \right) +  \sum_{k=1}^3 \partial_{x_k} \left\{ \left(
    \tfrac{1}{2} \rho \left| \bu \right|^{2} + \rho e + p \right) u_{k}
    \right\} = 0.
\nonumber
\end{align}
These equations are the transfer equations for moments \eqref{macroscopic quantites}, and they are known as Euler's gas dynamics equations. Observe that central moments of the distribution function, which represent non-convective macroscopic fluxes, have the form:
\beq \label{macroscopic fluxes5}
\lp \begin{matrix} 0_{k} \\ p_{ik} \\  q_{k} \end{matrix} \rp =  \int_{\mathbb{R}^3 \times \mathbb{R}_+} \lp \begin{matrix} m \, c_{k} \\ m \, c_{i} c_{k} \\    \frac{m}{2} \left( \left| \bc \right|^2 + I \right) c_{k} \end{matrix} \rp \, \hat{f}_{5} \, \vfi \, \md I \, \md \bv =
\lp \begin{matrix} 0_{k} \\ p \, \delta_{ik} \\  0_{k} \end{matrix} \rp.
\eeq
In other words, pressure tensor is diagonal, $p_{ik} = p \, \delta_{ik}$, where $p = (1/3) \sum_{i=1}^{3} p_{ii}$ is the equilibrium pressure, and the flux of internal energy (heat flux) $q_{k}$ vanishes. Moreover, equilibrium pressure $p$ is related to internal energy $e$ in the following way:
\begin{equation}\label{Eq_Pressure}
  p = \left( \alpha + \frac{5}{2} \right)^{-1} \rho e.
\end{equation}

\section{Polyatomic gases with dynamic pressure}

When the list of field variables \eqref{macroscopic quantites} is extended with non-conserved ones, we enter the field of non-equilibrium processes. The simplest one is characterized by a single non-equilibrium variable that can be interpreted as dynamic pressure. In the spirit of previous Section, we shall postpone the introduction of the dynamic pressure as a physical quantity, but rather start with purely mathematical application of MEP using 6 macroscopic fields. As an outcome, the non-equilibrium distribution function will be obtained.

\subsection{Non-equilibrium distribution function}

It was shown, in different approaches to non-equilibrium processes, that the simplest possible non-equilibrium model is the one which includes a single non-conserved field variable, apart from conserved ones \eqref{macroscopic quantites}. It was interpreted as internal variable in Meixner's approach, but naturally appears as dynamic pressure in extended thermodynamics of polyatomic gases \cite{ATRS_Meixner}. In application of MEP we shall introduce it, at first, as a trace of the second moment in momentum hierarchy---the momentum flux $P_{ij}$:
\beq \label{6field_moment}
  \sum_{i=1}^{3} P_{ii} = \sum_{i=1}^{3} \int_{\bR \times \bRp} m \, v_{i} v_{i} f \, \vfi \, \md I \, \md \bv.
\eeq
Corresponding central moment reads:
\beq \label{6field_central-moment}
  \sum_{i=1}^{3} p_{ii} = \sum_{i=1}^{3} \int_{\bR \times \bRp} m \, c_{i} c_{i} f \, \vfi \, \md I \, \md \bv.
\eeq
Our goal in the sequel is to show that $(1/3) \sum_{i=1}^{3} p_{ii}$ can be identified with equilibrium pressure $p$ only in local equilibrium, while the excess non-equilibrium pressure is the dynamic pressure $\Pi$.

To that end, consider the kinetic entropy \eqref{entropy} and the following variational problem which expresses the maximum entropy principle:
\begin{alignat}{3}\label{MEP: var problem}
\text{max}_f& \quad h  \nonumber\\
\text{s.t.}  & \lp \begin{array}{c}
    \rho \\
    0_i \\
    \sum_{i=1}^3 p_{ii} \\
    \rho e
  \end{array}\rp   =  \int_{\bR \times \bRp}
     \left(
  \begin{array}{c}
    m \\
   m\, c_{i} \\
   m\, \left| \bc \right|^2\\
   \frac{m}{2} \left| \bc \right|^2 + I
  \end{array}
  \right)\,
    f \, \vfi \, \md I \, \md\bc.
\end{alignat}
The solution of the problem for the choice $\vfi = I^{\alpha}$ is as follows.

\begin{proposition}
The  distribution function which maximizes the entropy \eqref{entropy} subject to constraints \eqref{MEP: var problem}, for the choice of the weighting function $\varphi(I) = I^{\alpha}$ has the form:
\begin{align}\label{solution MEP}
  \hat{f}_{6}(t, \bx, \bc + \bu, I) & = \frac{\rho}{m} \lp \frac{3 \rho}{2 \pi \sum_{i=1}^3 p_{ii}}
    \rp^{\frac{3}{2}} \lp \frac{\alpha+1}{m \, \lp e - \frac{1}{2 \rho} \sum_{i=1}^3 p_{ii} \rp} \rp^{\alpha+1} \frac{1}{\Gamma\left[ \alpha+1\right]}
  \\
  & \times \exp \left( -\frac{3 \rho}{2 \sum_{i=1}^3 p_{ii}} \left|  \bc \right|^2 - \frac{\alpha+1}{m \, \lp e - \frac{1}{2 \rho} \sum_{i=1}^3 p_{ii} \rp} I \right).
  \nonumber
\end{align}
\end{proposition}
The variational problem is solved by introducing the vector of multipliers
\beq\label{multipliers}
\lp \lambda^{(0)} \quad \lambda^{(1)}_1 \quad \lambda^{(1)}_2 \quad \lambda^{(1)}_3 \quad  \lambda^{(2)} \quad  \mu^{(2)} \rp^T.
\eeq
Then, the appropriate extended functional $\mathcal{L}$ reads
\begin{multline*}
\mathcal{L} = h - \lambda^{(0)}   \int_{\bR \times \bRp}  m  \,  f \, I^\alpha \, \md I \, \md \bc - \sum_{i=1}^3 \lambda^{(1)}_i \int_{\bR \times \bRp}  m \, c_i  \,  f \, I^\alpha \, \md I \, \md \bc \\ - \lambda^{(2)} \int_{\bR \times \bRp}  m \,  \left| \bc \right|^2 \,  f \, I^\alpha \, \md I \, \md \bc -  \mu^{(2)}   \int_{\bR \times \bRp} \lp \frac{m}{2} \left| \bc \right|^2 + I\rp \, f\, I^\alpha \, \md I\, \md\bc.
\end{multline*}
The Euler-Lagrange equation reduces to $\delta \mathcal{L}/\delta f =0$, which is satisfied if and only if
\beqs
- k \lp \log f +1 \rp - m \lp \lambda^{(0)} + \sum_{i=1}^3 \lambda^{(1)}_i c_i + \lambda^{(2)} \left| \bc \right|^2 \rp - \mu^{(2)} \lp \frac{m}{2} \left|  \bc \right|^2 + I \rp  =0.
\eeqs
Therefore, the solution $\hat{f}:= \hat{f}(t, \bx, \bv, I)$ to the variational problem \eqref{MEP: var problem} is
\begin{equation}\label{solution general MEP}
\hat{f}(t, \bx, \bc + \bu, I) = e^{-1 - \frac{m}{k} \lp \lambda^{(0)} + \sum_{i=1}^3 \lambda^{(1)}_i c_i + \lambda^{(2)} \left| \bc \right|^2 \rp - \frac{1}{k} \mu^{(2)} \lp \frac{m}{2} \left|  \bc \right|^2 + I \rp  }.
\end{equation}
Plugging this distribution function into the constraints of the problem \eqref{MEP: var problem}, we obtain a system of equations for the multipliers \eqref{multipliers} whose solution is
\begin{align*}
e^{- 1 - \frac{m}{k} \lambda^{(0)} }& = \frac{\rho}{m}\lp \frac{3 \rho}{  2 \pi \sum_{i=1}^3 p_{ii}}\rp^\frac{3}{2}  \lp \frac{\alpha+1}{m \, \lp e - \frac{1}{2 \rho} \sum_{i=1}^3 p_{ii} \rp} \rp^{\alpha+1} \frac{1}{\Gamma\left[ \alpha+1\right]},\\
\lambda_{i}^{(1)} & = 0, \qquad i=1,2,3,\\
\lambda^{(2)} & =  \frac{k}{2 \,m} \lp \frac{3 \rho}{\sum_{i=1}^3 p_{ii}} - \frac{\alpha+1}{ \lp e - \frac{1}{2 \rho} \sum_{i=1}^3 p_{ii} \rp} \rp, \\
\mu^{(2)} &= \frac{k \lp \alpha+1 \rp }{m \, \lp e - \frac{1}{2 \rho} \sum_{i=1}^3 p_{ii} \rp}.
\end{align*}
Insertion of the coefficients above into \eqref{solution general MEP} yields the non-equilibrium distribution function \eqref{solution MEP}.

\subsection{Field equations and source term}

Integration of the Boltzmann equation \eqref{BE} against $m, m \bv, m \left| \bv \right|^2, \frac{m}{2}  \left| \bv \right|^2 + I $ leads to the balance laws for mass density, momentum, trace of the momentum flux and energy density:
\begin{equation}\label{full system field eqs}
\begin{split}
  & \partial_{t} \rho + \sum_{k=1}^3 \partial_{x_k} (\rho u_{k}) = 0,
\\
  & \partial_{t} (\rho u_{i}) + \sum_{k=1}^3 \partial_{x_k}
    (\rho u_{i} u_{k} + p_{ik}) = 0,
\\
  & \partial_{t} \left( \rho \left| \bu \right|^2 + \sum_{i=1}^{3} p_{ii} \right)
      + \sum_{k=1}^3 \partial_{x_k} \left\{ \left( \rho \left| \bu \right|^2
      + \sum_{i=1}^{3} p_{ii} \right) u_{k} + 2 \sum_{i=1}^3 p_{ki} u_{i}
      +  \sum_{i=1}^3 p_{iik} \right\}  = \sum_{i=1}^3 \Pi_{ii}
\\
  & \partial_{t} \left( \tfrac{1}{2} \rho \left| \bu \right|^{2} +
    \rho e \right) +  \sum_{k=1}^3 \partial_{x_k} \left\{ \left(
    \tfrac{1}{2} \rho \left| \bu \right|^{2} + \rho e \right)
    u_{k} + \sum_{i=1}^3 p_{ki} u_{i} + q_{k} \right\} = 0,
\end{split}
\end{equation}
where the non-convective fluxes are defined as:
\begin{equation}\label{fluxes}
\left( \begin{matrix} 0_{k} \\ p_{ik} \\ \sum_{i=1}^3 p_{iik} \\ q_{k} \end{matrix} \right) = \int_{\bR \times \bRp} \left( \begin{matrix} m c_{k} \\ m c_i c_k \\ m \left| \bc \right|^2 c_k \\ \lp \frac{m}{2} \left| \bc \right|^2 + I \rp  c_{k} \end{matrix} \right) f \, \vfi \, \md I \, \md \bv,
\end{equation}
and the production term is given with:
\begin{equation}\label{production term}
\sum_{i=1}^3 \Pi_{ii} = \int_{\bR \times \bRp} m \inzv^2 Q(f,f)(\bv, I) \, \vfi \, \md I \, \md \bv.
\end{equation}
Equations \eqref{full system field eqs}$_{1-3}$ belong to the momentum hierarchy, while \eqref{full system field eqs}$_{4}$ is the representative of the energy hierarchy of the moment equations.

The system \eqref{full system field eqs} is not closed, since the non-convective fluxes \eqref{fluxes} and the production term \eqref{production term} are not determined. Closure is achieved through the application of MEP, i.e. calculation of undetermined fluxes and source terms using the entropy maximizer, $\hat{f}_{6}$ in this case.



To calculate the non-convective fluxes, it is sufficient to fix the distribution function and choose the weight function. For $f = \hat{f}_{6}$ and $\vfi = I^{\alpha}$, $\alpha > -1$, the non-convective fluxes read:
\begin{equation}\label{fluxes6}
  p_{ik} = \delta_{ik} \frac{1}{3}\sum_{j=1}^3 p_{jj}; \quad
  \sum_{i=1}^3 p_{iik} = 0; \quad q_{k} =  0.
\end{equation}
Calculation of the production term is more delicate, since one has to choose the collision cross section $\mathcal{B}$, i.e. the model of molecular interaction. This choice strongly influences the structure of production term. In the sequel, we shall analyze two possible choices of the collision cross section.

\begin{proposition}
  Assume that the non-equilibrium state of the polyatomic gas is described by the distribution function \eqref{solution MEP}, in accordance with the weight function $\vfi = I^{\alpha}$, $\alpha > -1$.
  \begin{itemize}
    \item[(a)] If the collision cross section is chosen as:
      \begin{equation}\label{cross-secction1}
        \mathcal{B}(\bv, \bv_*, I, I_*, r, R, \bom) = K R^s \left| \bv-\bv_* \right|^{2s},
      \end{equation}
    where $K$ is an appropriate dimensional constant, and $s\in \mathbb{R}$ a parameter satisfying $s>-\frac{3}{2}$, corresponding production term reads:
      \begin{multline}\label{production term explicit special case}
        \sum_{i=1}^3 \Pi_{ii} = K \, \frac{\rho^2}{m}\, \pi^{1/2} \, \frac{2^{2s+6} \, \Gamma\left[ s + \frac{3}{2} \right] }{\lp 2 s+ {5} \rp \lp 2 s+ {7} \rp}  \lp \frac{1} {3\rho} \sum_{i=1}^3 p_{ii} \rp^{s} \frac{1}{\Gamma\left[ \alpha + 1 \right]^2} \lp  \frac{\alpha+1 }{m \, \lp e - \frac{1}{2 \rho} \sum_{i=1}^3 p_{ii} \rp } \rp^{2 \alpha}
        \\
        \times \lp - \frac{1} {3\rho} \sum_{i=1}^3 p_{ii} + \frac{1}{\alpha+1} \lp e - \frac{1}{2 \rho} \sum_{i=1}^3 p_{ii} \rp \rp.
      \end{multline}
    \item[(b)] If the collision cross section is chosen as:
      \begin{equation}\label{cross-secction2}
        \mathcal{B}_{G}(\bv, \bv_*, I, I_*, r, R, \bom)=K_{G} R^s \left| \bv - \bv_*\right|^{2s}
        (I+I_*)^{\beta}(1-R)^{\beta}\left| \bG \right|^{2q},
      \end{equation}
    where $\bG = \bv + \bv_*$, and $s,q>-3/2$, $\beta>-2$, corresponding production term reads:
      \begin{multline}\label{production term explicit general case}
        \sum_{i=1}^3 \Pi_{ii}=  K_{G} \frac{\rho^2}{m}  2^{2s+4} \Gamma\left[q+\frac{3}{2}\right] \frac{\Gamma\left[s+\frac{5}{2}\right]\Gamma\left[s+\frac{3}{2}\right]
        \Gamma[\beta+2]\Gamma[\beta+3]}{\Gamma\left[s+\beta+\frac{9}{2}\right]}
        \\
        \times \left(  \frac{1}{3 \rho} \sum_{i=1}^3 p_{ii} \right)^{s+q} \frac{1}{\Gamma[\alpha+1]^2} \left( \frac{(\alpha+1)}{m\left(e-\frac{1}{2\rho} \sum_{i=1}^3 p_{ii}\right)} \right)^{2\alpha-\beta}
        \\
        \times  \lp - \frac{1} {3\rho} \sum_{i=1}^3 p_{ii} + \frac{1}{\alpha+1} \lp e - \frac{1}{2 \rho} \sum_{i=1}^3 p_{ii} \rp \rp. 
      \end{multline}
  \end{itemize}
\end{proposition}
Proof of the Proposition 5 is postponed to Appendix.

\paragraph{Remark on the production term.} It is obvious that the cross section \eqref{cross-secction1} is a special case of the cross section \eqref{cross-secction2}, obtained for $q = 0$, $\beta = 0$. The reason for the analysis of both cases is the following. Case (a) presents one of the standard choices for the cross section, here extended with parameter $R$ that describes the distribution of total energy between the kinetic energy of the molecule and microscopic internal energy. In the case (b), the cross section $\mathcal{B}_{G}$ comprises also the total microscopic internal energy, $I + I_*$, as well as the average peculiar velocity of the colliding molecules, $|\bG|$. This considerably rich structure contains 3 parameters, $s$, $q$ and $\beta$, and provides the possibility for better matching with the results obtained by other approaches. In particular, it will be shown that \eqref{production term explicit general case} can be related to the production term obtained in extended thermodynamics, whereas \eqref{production term explicit special case} can not. However, there is a price for this result, as will be explained in the next remark.

\paragraph{Remark on Galilean invariance.} Mathematical theory of the Boltzmann equations requires that the collision cross section is microreversible, as already mentioned in Section 2. On the other hand, collisional physics imposes also the condition of invariance with respect to Galilean transformations. The same requirement of invariance is imposed also on the macroscopic equations. This is inevitably true for collision transformation \eqref{microscopic CL}, as well as for the ``standard'' cross section \eqref{cross-secction1}. However, generalized cross section \eqref{cross-secction2} is microreversible, but it is not Galilean invariant due to the presence of $\bG$. Is it physically correct to use the cross section \eqref{cross-secction2} since it is merely microreversible? Interestingly enough, although \eqref{cross-secction2} is not Galilean invariant, it yields the Galilean invariant production term \eqref{production term explicit general case}. In other words, integration over the velocity space ``wipes-out'' the non-invariance at microscopic level. From the point of view of macroscopic dynamics, this leaves greater freedom in the choice of the cross-section because not all of its micro properties are transferred to the macro level.

\subsection{Entropy density and entropy production}

In the macroscopic, or phenomenological approach, the entropy density and the entropy production come out from the compatibility of the moment equations with the entropy balance law, while unknown non-convective fluxes and production terms secure the validity of the entropy inequality for all thermodynamic processes. MEP sheds a bit different light on the same quantities. Once we found the entropy maximizer, the entropy density is uniquely determined.

For polyatomic gases in local equilibrium (described with 5 macroscopic fields), entropy maximizer \eqref{solution MEP eq} yields the following entropy density \eqref{entropy}:
\begin{align} \label{entropy_5}
  \hat{h}_{5} & = h\left( \hat{f}_{5} \right)
  \\
  & = - k \frac{\rho}{m} \left(
    \log \left\{ \frac{\rho}{m \Gamma[\alpha+1]} \left( \left( \alpha + \frac{5}{2} \right)
    \frac{1}{ m e} \right)^{\alpha+1}
    \left( \left( \alpha + \frac{5}{2} \right) \frac{1}{2 \pi e} \right)^{3/2} \right\}
    - \left( \alpha + \frac{5}{2} \right) \right).
  \nonumber
\end{align}
For polyatomic gases described with 6 fields, the entropy density is determined by the maximizer \eqref{solution MEP}:
\begin{align} \label{entropy_6}
  \hat{h}_{6} & = h\left( \hat{f}_{6} \right)
  \\
  & = - k \frac{\rho}{m} \left(
    \log \left\{ \frac{\rho}{m \Gamma[\alpha+1]}
    \left( \frac{\alpha +1 }{m\left( e - \frac{1}{2\rho} \sum_{i=1}^3 p_{ii}\right)} \right)^{\alpha+1}
    \left( \frac{3 \rho}{2 \pi \sum_{i=1}^3 p_{ii}} \right)^{3/2} \right\}
    - \left( \alpha + \frac{5}{2} \right) \right).
  \nonumber
\end{align}

In contrast to entropy density, entropy production \eqref{entropy production} does not depend solely on the velocity distribution function, but on the cross section as well. In the most general case studied here, i.e for the entropy maximizer \eqref{solution MEP} and the cross section \eqref{cross-secction2}, the entropy production \eqref{entropy production} reads:
\begin{align} \label{entropy_production-gen}
  D(\hat{f}_{6}, \mathcal{B}_{G}) & = - K \frac{\rho^2}{m^3} \frac{1}{\Gamma[\alpha+1]^2} 2^{2s+3}
    \Gamma\left[q+\frac{3}{2}\right] \frac{\Gamma\left[s+\frac{5}{2}\right] \Gamma\left[s+\frac{3}{2}\right] \Gamma[\beta+2]\Gamma[\beta+3]}{\Gamma\left[s+\beta+\frac{9}{2}\right]}
  \\
    & \times \left( \frac{(\alpha+1)}{m\left(e-\frac{1}{2\rho} \sum_{i=1}^3 p_{ii}\right)} \right)^{2\alpha-1-\beta} \left( \frac{3\rho}{\sum_{i=1}^3 p_{ii}}  \right)^{-s-1-q}
    \left( - \frac{3\rho}{\sum_{i=1}^3 p_{ii}} + \frac{(\alpha+1)}{\left(e-\frac{1}{2\rho}
    \sum_{i=1}^3 p_{ii}\right)} \right)^2.
  \nonumber
\end{align}

\subsection{Dynamic pressure}

As already indicated, we did not introduce the dynamic pressure as non-equilibrium field variable from the outset. Our aim was to close the 6 moments model for polyatomic gases using MEP, and then to recognize that dynamic pressure appears as a variable that substantially describes the non-equilibrium part of the entropy. To that end, we shall introduce the simplified notation, $h \equiv h(\hat{f}_{6})$ and $h_{E} \equiv h(\hat{f}_{5})$, and define the non-equilibrium part of the entropy $\bar{\kappa} := h - h_E$ that in the 6 moments model takes the following form:
\begin{equation} \label{kappa6}
  \bar{\kappa}(\rho, e, \sum_{i=1}^3 p_{ii}) = -k \frac{\rho}{m}  \log \left\{ \lp \frac{\lp \alpha +1  \rp m e}{\lp \alpha + \frac{5}{2} \rp m \lp e-\frac{1}{2\rho} \sum_{i=1}^3 p_{ii} \rp } \rp^{\alpha+1}  \lp \frac{3 \rho e}{\lp \alpha + \frac{5}{2} \rp \sum_{i=1}^3 p_{ii}} \rp^{3/2} \right\}.
\end{equation}

As $h$ reaches the equilibrium value $h_{E}$, $\bar{\kappa}$ vanishes. It can be easily shown that $\bar{\kappa} = 0$ when:
\begin{equation}\label{pii_Eq}
  \sum_{i=1}^{3} p_{ii} = \frac{3 \rho e}{\alpha + \frac{5}{2}} = 3 p,
\end{equation}
i.e. trace of the pressure tensor in equilibrium is related to thermodynamic pressure, as expected. Therefore, we may introduce the scalar field variable $\Pi$, \emph{dynamic pressure}, which determines the non-equilibrium part of the trace of the pressure tensor:
\begin{equation}\label{dynamic pressure-def}
  \Pi:=\frac{1}{3} \sum_{i=1}^3 p_{ii} - p.
\end{equation}
Then $\sum_{i=1}^3 p_{ii} = 3 \Pi +3 \rho e \lp \alpha + \frac{5}{2} \rp^{-1}$, and \eqref{kappa6} can be written using ``physical'' variables:
\begin{align} \label{kappa6-Pi}
  \kappa(\rho,e,\Pi) & :=
    \bar{\kappa}\left(\rho, e,  3 \Pi +3 \rho e \lp \alpha + \frac{5}{2} \rp^{-1}\right)
  \\
  & = k \frac{\rho}{m}  \log \left\{ \lp 1-\frac{3}{2}
    \frac{\lp \alpha + \frac{5}{2}\rp}{(\alpha+1)\rho e}\Pi \rp^{\alpha+1}  \lp 1 +  \lp \alpha + \frac{5}{2}\rp \frac{1}{\rho e} \Pi \rp^{3/2} \right\},
  \nonumber
\end{align}
such that $\kappa(\rho, e, 0) = 0$.

Once the dynamic pressure is introduced, all the macroscopic equations may be rewritten in terms of the new set of macroscopic variables $(\rho, \mathbf{u}, e, \Pi)$. Field equations \eqref{full system field eqs} will have the following form:
\begin{equation}\label{full system field eqs-Pi}
\begin{split}
  & \partial_{t} \rho + \sum_{k=1}^3 \partial_{x_k} (\rho u_{k}) = 0,
\\
  & \partial_{t} (\rho u_{i}) + \sum_{k=1}^3 \partial_{x_k}
    (\rho u_{i} u_{k} + (p + \Pi) \delta_{ik}) = 0,
\\
  & \partial_{t} \left( \rho \left| \bu \right|^2 + 3 (p + \Pi) \right)
      + \sum_{k=1}^3 \partial_{x_k} \left\{ \left( \rho \left| \bu \right|^2
      + 5 (p + \Pi) \right) u_{k} \right\}  = \sum_{i=1}^3 \Pi_{ii},
\\
  & \partial_{t} \left( \tfrac{1}{2} \rho \left| \bu \right|^{2} +
    \rho e \right) +  \sum_{k=1}^3 \partial_{x_k} \left\{ \left(
    \tfrac{1}{2} \rho \left| \bu \right|^{2} + \rho e + p + \Pi \right) u_{k} \right\} = 0.
\end{split}
\end{equation}
where equilibrium pressure is determined by \eqref{Eq_Pressure}. The source term \eqref{production term explicit special case} reads:
\begin{multline}\label{production term explicit special case-Pi}
  \sum_{i=1}^3 \Pi_{ii} = - K \, \frac{\rho^2}{m^{2 \alpha + 1}}\, \pi^{1/2} \,
    \frac{2^{2s+4}}{\lp s + \frac{5}{2} \rp \lp s + \frac{7}{2} \rp}
    \frac{\Gamma\left[ s + \frac{3}{2} \right]}{\Gamma\left[ \alpha + 1 \right]^2}
    \frac{\alpha + \frac{5}{2}}{\alpha + 1} \frac{\Pi}{\rho}
  \\
    \times \lp \frac{e}{\alpha + \frac{5}{2}} + \frac{\Pi}{\rho} \rp^{s}
    \lp \frac{e}{\alpha + \frac{5}{2}} - \frac{3}{2 (\alpha + 1)} \frac{\Pi}{\rho} \rp^{- 2 \alpha},
\end{multline}
while the source term \eqref{production term explicit general case} reads:
\begin{multline}\label{production term explicit general case-Pi}
  \sum_{i=1}^3 \Pi_{ii} = - K_{G} \, \frac{\rho^2}{m^{2 \alpha - \beta + 1}}\, 2^{2s+4} \,
    \frac{\Gamma\left[ q + \frac{3}{2} \right] \Gamma\left[ s + \frac{5}{2} \right]
      \Gamma\left[ s + \frac{3}{2} \right] \Gamma\left[ \beta + 2 \right]
      \Gamma\left[ \beta + 3 \right]}{\Gamma\left[ s + \beta + \frac{9}{2} \right]
      \Gamma\left[ \alpha + 1 \right]^2}
    \frac{\alpha + \frac{5}{2}}{\alpha + 1} \frac{\Pi}{\rho}
  \\
    \times \lp \frac{e}{\alpha + \frac{5}{2}} + \frac{\Pi}{\rho} \rp^{s + q}
    \lp \frac{e}{\alpha + \frac{5}{2}} - \frac{3}{2 (\alpha + 1)} \frac{\Pi}{\rho} \rp^{- (2 \alpha - \beta)}.
\end{multline}

\subsection{Comparison with BGK model and ET near equilibrium}

Source term, either in the form \eqref{production term explicit special case-Pi}, or in the form \eqref{production term explicit general case-Pi}, presents a novel results in comparison with the recent studies of polyatomic gases within ET. Therefore, it should be compared with them in order to recognize its peculiarities.

Consider the linearized form of the source term, where linearization is assumed only with respect to dynamic pressure $\Pi$:
\begin{align}
  \sum_{i=1}^3 \Pi_{ii}^{lin} & = - K \, \frac{\rho^2}{m^{2 \alpha + 1}} \, C(\alpha, s) \,
    \lp \frac{e}{\alpha + \frac{5}{2}} \rp^{s - 2 \alpha} \frac{\Pi}{\rho},
  \label{production term explicit special case-BGK} \\
  \sum_{i=1}^3 \Pi_{ii}^{lin} & = - K_{G} \, \frac{\rho^2}{m^{2 \alpha - \beta + 1}} \,
    C_{G}(\alpha, s, \beta, q) \,
    \lp \frac{e}{\alpha + \frac{5}{2}} \rp^{s - 2 \alpha + q + \beta}
    \frac{\Pi}{\rho},
  \label{production term explicit general case-BGK}
\end{align}
where the $C(\alpha, s)$ and $C_{G}(\alpha, s, \beta, q)$ stand for:
\begin{align}
  C(\alpha, s) & = \pi^{1/2} \,
    \frac{2^{2s+4}}{\lp s + \frac{5}{2} \rp \lp s + \frac{7}{2} \rp}
    \frac{\Gamma\left[ s + \frac{3}{2} \right]}{\Gamma\left[ \alpha + 1 \right]^2}
    \frac{\alpha + \frac{5}{2}}{\alpha + 1},
  \label{production term special case-C} \\
  C_{G}(\alpha, s, \beta, q) & = 2^{2s+4} \,
    \frac{\Gamma\left[ q + \frac{3}{2} \right] \Gamma\left[ s + \frac{5}{2} \right]
      \Gamma\left[ s + \frac{3}{2} \right] \Gamma\left[ \beta + 2 \right]
      \Gamma\left[ \beta + 3 \right]}{\Gamma\left[ s + \beta + \frac{9}{2} \right]
      \Gamma\left[ \alpha + 1 \right]^2} \frac{\alpha + \frac{5}{2}}{\alpha + 1}.
  \label{production term general case-C}
\end{align}
Actually, linearized form of the source term was the first one considered in ET, and the only one considered so far in the MEP for polyatomic gases with 6 fields. It is akin to the results that are obtained the use of BGK model equation.

To appreciate the difference between the source terms \eqref{production term explicit special case-BGK}-\eqref{production term explicit general case-BGK} and the ones presented in ET, assume that they can be expressed in the form $\sum_{i=1}^{3} \Pi_{ii}^{lin} = - 3 \Pi/\tau_{\Pi}$, where $\tau_{\Pi}$ denotes the relaxation time for dynamic pressure. In our case, relaxation times have the following form:
\begin{align}
  \tau_{\Pi} & = 3 \left[ K \, \frac{\rho^2}{m^{2 \alpha + 1}} \, C(\alpha, s) \,
    \lp \frac{e}{\alpha + \frac{5}{2}} \rp^{s - 2 \alpha} \right]^{-1},
  \label{relaxation time - special case} \\
  \tau_{\Pi} & = 3 \left[ K_{G} \, \frac{\rho^2}{m^{2 \alpha - \beta + 1}} \,
    C_{G}(\alpha, s, \beta, q) \,
    \lp \frac{e}{\alpha + \frac{5}{2}} \rp^{s - 2 \alpha + q + \beta} \right]^{-1},
  \label{relaxation time - general case}
\end{align}
corresponding to \eqref{production term explicit special case-BGK} and \eqref{production term explicit general case-BGK}, respectively. Apart from parameter $\alpha$ which determines internal degrees of freedom, there appear one more parameter $s$ in \eqref{relaxation time - special case}, or additional two, $\beta$ and $q$, in \eqref{relaxation time - general case}. They bring the flexibility in our model, and allow matching with the data obtained experimentally, or by some other theoretical approach. In contrast to this result, ET with linear closure does not bring any information about the value of $\tau_{\Pi}$, except its positivity. This is not peculiarity of ET, but common for all macroscopic theories. In this case, one must take the information `from the outside' and incorporate it in the analysis of particular processes.

\subsection{Comparison with ET far from equilibrium}

As already mentioned in the Introduction, polyatomic gas with 6 fields admits non-linear closure within ET. As a matter of fact, source term is expressed in the form:
\begin{equation}\label{production term ET}
  \sum_{i=1}^{3} \Pi_{ii} = \bar{\alpha}(\rho,e) \partial_{\Pi} \kappa,
\end{equation}
where $\bar{\alpha}(\rho,e) > 0$ is a phenomenological function, and $\kappa$ is non-equilibrium part of the entropy. In ET, should be the solution of a quasi-linear PDE (eq. (25) in \cite{ARST_NLET6}), that can be obtained using the method of characteristics. In the end, the non-linear source term of ET can be expressed in our notation:
\begin{equation}\label{production term ET-our}
  \sum_{i=1}^{3} \Pi_{ii} = - \frac{3}{\tau_{\Pi}}
    \frac{\Pi}{\left( 1 + \left( \alpha + \frac{5}{2} \right) \frac{\Pi}{\rho e} \right)
    \left( 1 - \frac{3}{2(\alpha + 1)} \left( \alpha + \frac{5}{2} \right) \frac{\Pi}{\rho e} \right)}.
\end{equation}

It is obvious, at first sight, that source terms \eqref{production term explicit special case-Pi} and \eqref{production term explicit general case-Pi} have more general structure than \eqref{production term ET-our}. Therefore, one may expect that they can be reduced to the latter form for special values of parameters. First, it can be shown that non-equilibrium part of the entropy $\kappa(\rho,e,\Pi)$ satisfies PDE (25) from \cite{ARST_NLET6}. However, surprisingly enough, source term \eqref{production term explicit special case-Pi}, obtained using standard cross section, cannot be reduced to non-linear source term of ET. This fact motivated the analysis of non-standard cross sections that will lead to equivalent macroscopic source term. It can be shown that \eqref{production term ET-our} can be obtained from \eqref{production term explicit general case} for the specific choice of parameters:
\begin{equation}\label{ET_Compatibility}
  \beta = 2 \alpha - 1; \quad q = - (s + 1),
\end{equation}
and in this case:
\begin{equation*}
\bar{\alpha}  (\rho,e) = \frac{1}{3} \frac{K}{k} \frac{\rho^2}{m} \frac{1}{\Gamma[\alpha+1]^2} 2^{2s+5} \frac{\Gamma\left[s+\frac{5}{2}\right]\Gamma\left[s+\frac{3}{2}\right] \Gamma[-s+\frac{1}{2}]\Gamma[2\alpha+1]\Gamma[2 \alpha+2]}{\Gamma\left[s+2\alpha+\frac{7}{2}\right]}.
\end{equation*}

\section{Shock structure in polyatomic gases}
\newcommand{\D}{\mathrm{d}}

Shock structure is a continuous solution of the dissipative system of field equations that connects two equilibrium states. Its simplest form can be represented as a traveling wave, moving at a constant speed $\sigma$, with steep gradients of state variables in the neighborhood of a plane orthogonal to the line of propagation. Continuity of the solution is expected due to the presence of dissipative mechanism that smears out the jumps of field variables---the shock waves. However, in hyperbolic systems of balance laws there could also appear the solution that contains a sub-shock. As it was proved in \cite{boillat1998shock}, such a solution appears when the speed of the traveling wave is greater than the highest characteristic speed of the system in upstream equilibrium.

In \cite{RET_Poly} the shock structure in polyatomic gases was studied, and three types of solution were distinguished: (A) symmetric solutions that occur for low values of upstream Mach number; (B) asymmetric solutions---still continuous solutions for a bit higher values of upstream Mach number; (C) solutions with sub-shock. Our aim is to analyze the shock structure in polyatomic gases, taking into account specific structure of the source term, and to perform the parametric study of the solutions.

Consider the system of governing equations \eqref{full system field eqs-Pi} and assume the one-dimensional macroscopic motion of the polyatomic gas in $x$-direction, in which the velocity field reads $\mathbf{u} = (u, 0, 0)$. Also, assume that solution has the form of a traveling wave, so that all the field variables depend on a single independent variable $\xi = x - \sigma t$. These assumptions reduce the system \eqref{full system field eqs-Pi} to the system of ordinary differential equations---the shock structure equations:
\begin{equation}\label{equ:6fieldsShock}
\begin{split}
  & \partial_{\xi} (\rho \bar{u}) = 0 ,
\\
  & \partial_{\xi}
    (\rho \bar{u}^2 + p + \Pi) = 0,
\\
  &  \partial_{\xi} \left\{ \left( \rho \bar{u}^2
      + 5 (p + \Pi) \right) \bar{u} \right\}  = \sum_{i=1}^{3} \Pi_{ii},
\\
  & \partial_{\xi} \left\{ \left(
    \tfrac{1}{2} \rho \bar{u}^{2} + \rho e + p + \Pi \right) \right\} = 0
\end{split}
\end{equation}
where $\bar{u} = u - \sigma$. We shall also assume that the gas is ideal and obeys usual thermal and caloric equations of state:
\begin{equation}\label{EqState}
  p = \rho \frac{k}{m} T; \quad e = \left( \alpha + \frac{5}{2} \right) \frac{k}{m} T,
\end{equation}
where $T$ is absolute temperature.

The shock structure problem will be analyzed in dimensionless form. The following scaled variables will be used:
\begin{equation*}
\hat{\xi} = \frac{\xi}{L}, \hspace{0.3cm} \hat{\rho} = \frac{\rho}{\rho_0} \hspace{0.3cm} \hat{u} = \frac{\bar{u}}{c_0}, \hspace{0.3cm} \hat{T} = \frac{T}{T_0}, \hspace{0.3cm} \hat{p} = \frac{p}{\rho_0 \tfrac{k}{m}T_0}, \hspace{0.3cm} \hat{e} = \frac{e}{\left(\alpha + \tfrac{5}{2}\right)\tfrac{k}{m}T_0}, \hspace{0.3cm} \hat{\Pi} = \frac{\Pi}{\rho_0 \tfrac{k}{m}T_0},
\end{equation*}
where:
\begin{equation*}
  c_{0} = \sqrt{\frac{7 + 2 \alpha}{5 + 2 \alpha} \frac{k}{m} T_{0}}, \quad
  L = c_{0} \tau_{\Pi 0}.
\end{equation*}
Subscript $0$ indicates the upstream equilibrium state (in front of the shock), $c_{0}$ is the speed of sound in upstream equilibrium and $\tau_{\Pi 0}$ is the relaxation time for dynamic pressure in the same state. Reference relaxation time will be used either in the form \eqref{relaxation time - special case}, or in the form \eqref{relaxation time - general case}, depending on the source term that will be used in the analysis. Dimensionless shock structure equations have the following form (hats are dropped for simplicity):
\begin{equation} \label{DLessShockEquations}
\begin{split}
& \frac{\D}{\D \xi}\left(\rho u \right) = 0,
\\
& \frac{\D}{\D \xi}\left( \rho u^2 + \frac{5 + 2 \alpha}{7 + 2 \alpha} ( \rho T + \Pi ) \right) = 0,
\\
& \frac{\D}{\D \xi} \left\{ \rho u^3 + 5 \left( \frac{5 + 2 \alpha}{7 + 2 \alpha}
  ( \rho T + \Pi ) \right) \hat{u} \right\} = - 3 \, \frac{5 + 2 \alpha}{7 + 2 \alpha} \, \rho \Pi
  \left( T + \frac{\Pi}{\rho} \right)^{s^{\ast}}
  \left( T + \frac{3}{2 (\alpha + 1)} \frac{\Pi}{\rho} \right)^{- 2 \alpha^{\ast}},
\\
& \frac{\D}{\D \xi} \left\{ \left( \frac{1}{2} \rho u^2 + \left(\alpha + \frac{5}{2}\right) \rho T
  + \frac{5 + 2 \alpha}{7 + 2 \alpha} \Pi \right) u \right\} = 0.
\end{split}
\end{equation}
In \eqref{DLessShockEquations} $s^{\ast} = s$ and $\alpha^{\ast} = \alpha$ when source term is taken as \eqref{production term explicit special case-Pi}, while $s^{\ast} = s + q$ and $\alpha^{\ast} = \alpha - \beta/2$ for the source term \eqref{production term explicit general case-Pi}.

\subsection{Continuous shock structure}

First part of the analysis will be devoted to the continuous shock structure, i.e. shock structure without a sub-shock. Such a solution asymptotically connects the equilibrium states, upstream $(0)$ and downstream $(1)$, that are solutions of Rankine-Hugoniot equations for an equilibrium subsystem---Euler equations. In the state space these solutions represent heteroclinic orbits. Dimensionless state variables have the following values in equilibrium states:
\begin{equation}\label{RH-Euler}
  \left(
    \begin{array}{c}
      \rho_{0} \\
      u_{0} \\
      T_{0} \\
      \Pi_{0} \\
    \end{array}
  \right)
  =
  \left(
    \begin{array}{c}
      1 \\
      M_{0} \\
      1 \\
      0 \\
    \end{array}
  \right),
  \quad
  \left(
    \begin{array}{c}
      \rho_{1} \\
      u_{1} \\
      T_{1} \\
      \Pi_{1} \\
    \end{array}
  \right)
  =
  \left(
    \begin{array}{c}
      \frac{2(3 + \alpha)M_{0}^{2}}{5 + 2\alpha + M_{0}^{2}} \\
      \frac{5 + 2\alpha + M_{0}^{2}}{2(3 + \alpha)M_{0}} \\
      \frac{(7 + 2\alpha)M_{0}^{4} + (34 + 24 \alpha + 4 \alpha^{2})M_{0}^{2} - (5 + 2\alpha)}
        {4(3 + \alpha)^{2}M_{0}^{2}} \\
      0 \\
    \end{array}
  \right).
\end{equation}
In \eqref{RH-Euler}, $M_{0} = \bar{u}_{0}/c_{0}$ is Mach number in upstream equilibrium. Continuous shock structure is physically admissible if $M_{0} > 1$. Numerical solution of the shock structure equations \eqref{DLessShockEquations} with boundary data \eqref{RH-Euler} is obtained using the initial value algorithm, in the spirit of computations given in \cite{Simic-StabBif,Simic-Shock,damir2014}. All the profiles for $\rho$, $u$ and $T$ (not for $\Pi$) will be given in ``normalized'' form---values of the state variables will change between $0$ and $1$.

Careful analysis of symmetric and asymmetric continuous shock profiles was given in \cite{RET_Poly}. Our focus will be on the dependence of shock profiles on the values of parameters. First, we shall analyze the shock structure using the ``standard'' source term \eqref{production term explicit special case-Pi}.

\begin{figure}[h]
    \centering
    \begin{subfigure}[b]{0.38\textwidth}
        \includegraphics[width=\textwidth]{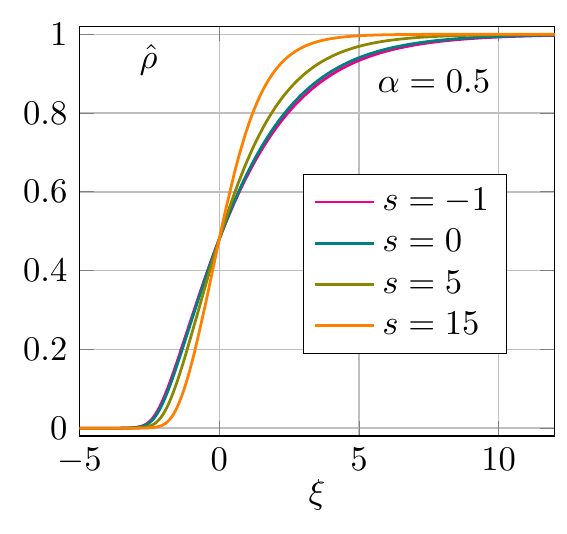}
        \label{fig:rho_param_1}
    \end{subfigure}
    ~ 
    \begin{subfigure}[b]{0.38\textwidth}
        \includegraphics[width=\textwidth]{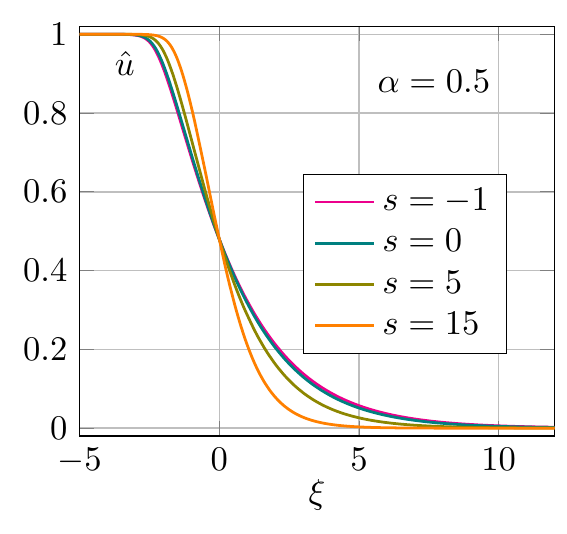}
        \label{fig:u_param_1}
    \end{subfigure}
    ~ 
    \begin{subfigure}[b]{0.38\textwidth}
        \includegraphics[width=\textwidth]{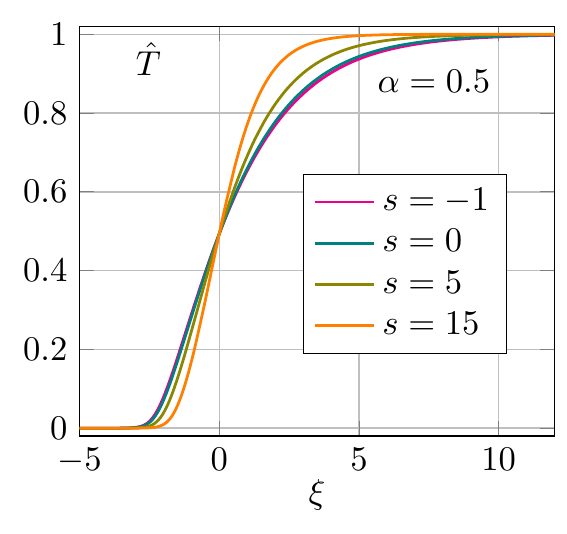}
        \label{fig:T_param_1}
    \end{subfigure}
    ~ 
    \begin{subfigure}[b]{0.38\textwidth}
        \includegraphics[width=\textwidth]{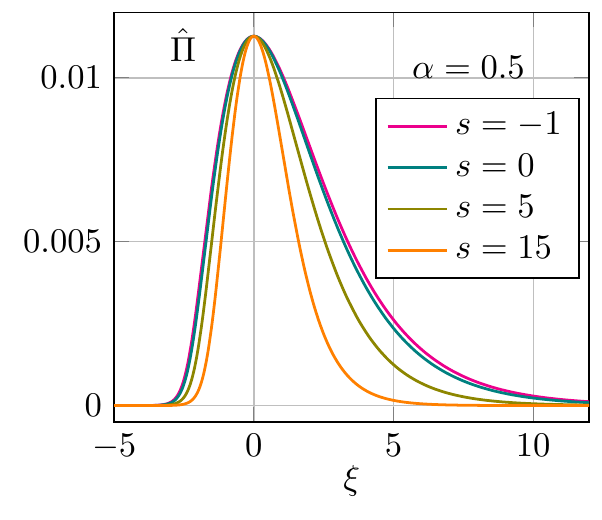}
        \label{fig:P_param_1}
    \end{subfigure}
    \caption{Shock profiles: $M_0=1.1$, $\alpha = 0.5$}\label{fig:param_1}
\end{figure}

Figure \ref{fig:param_1} shows the shock profiles of field variables for fixed $\alpha = 0.5$, corresponding to $\gamma = 4/3$, and $M_{0} = 1.1$, and several values of parameter $s$. It is obvious that the increase of $s$ leads to steeper profiles of state variables. This tendency remains the same also for different admissible values of $\alpha$ and $M_{0}$.

\begin{figure}[h]
    \centering
    \begin{subfigure}[b]{0.38\textwidth}
        \includegraphics[width=\textwidth]{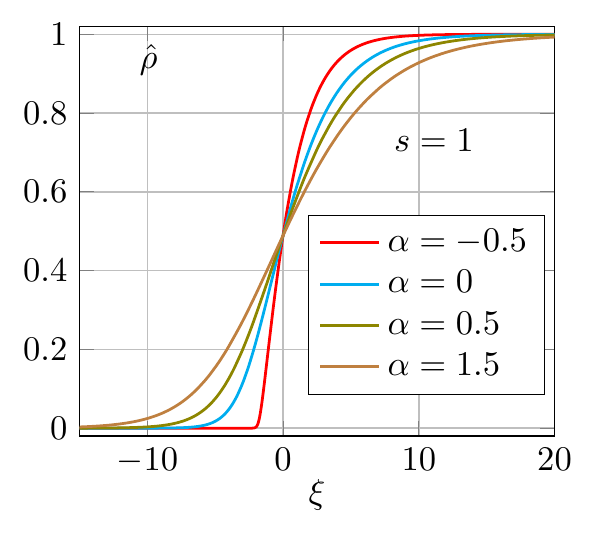}
        \label{fig:rho_param_2}
    \end{subfigure}
    ~ 
    \begin{subfigure}[b]{0.38\textwidth}
        \includegraphics[width=\textwidth]{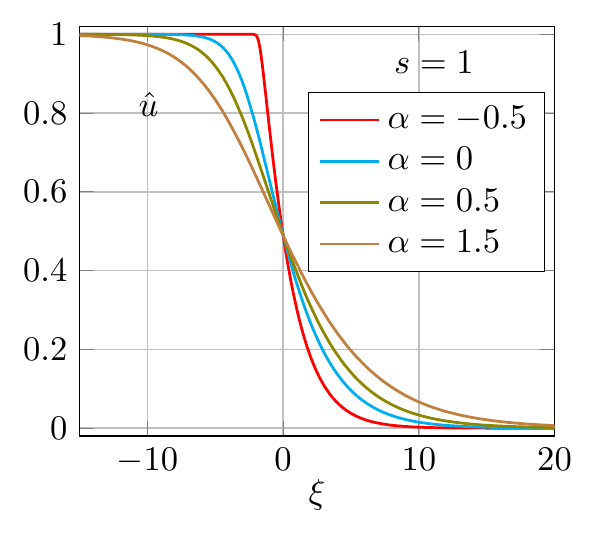}
        \label{fig:rho_param_2}
    \end{subfigure}
    ~ 
    \begin{subfigure}[b]{0.38\textwidth}
        \includegraphics[width=\textwidth]{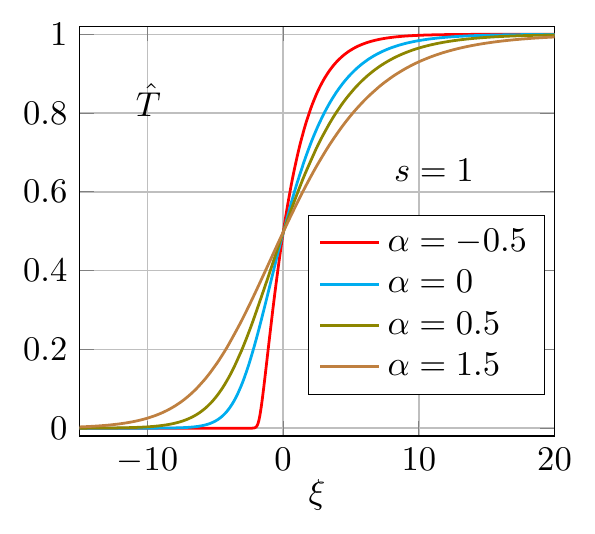}
        \label{fig:T_param_2}
    \end{subfigure}
    ~ 
    \begin{subfigure}[b]{0.38\textwidth}
        \includegraphics[width=\textwidth]{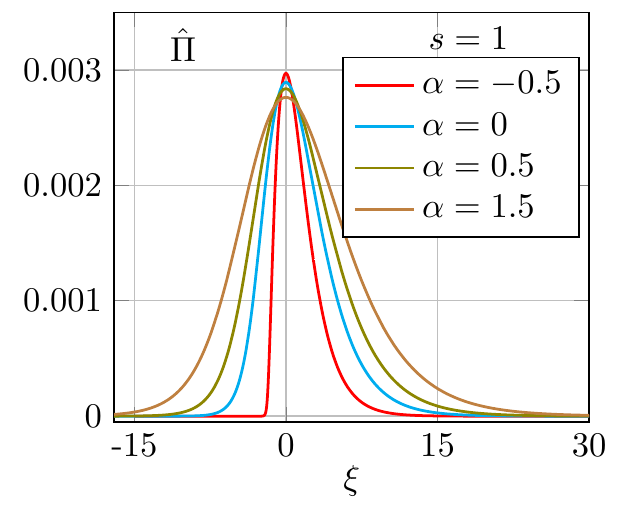}
        \label{fig:P_param_2}
    \end{subfigure}
    \caption{Shock profiles: $M_0=1.05$, $s = 1$}\label{fig:param_2}
\end{figure}

Figure \ref{fig:param_2} shows the shock profiles for fixed $s = 1.0$ and $M_{0} = 1.05$, and several values of parameter $\alpha$. To motivate the choice of the values note that $\alpha > -1$, and at $\alpha = -1$ there appears singularity in the model. Moreover, if admissible, $\alpha = -1$ would have correspond to the case of a monatomic gas. Therefore, we chose $\alpha = - 0.5$ to mimic the monatomic behaviour, and $\alpha = 1.5$ to mimic the polyatomic gas with excited internal degrees of freedom (e.g. vibration). Other two values correspond to polyatomic gases with two and three atoms. It is interesting to see that the increase of $\alpha$ implies smaller gradients of state variables and wider profiles. Therefore, more degrees of freedom require more space for transfer from one stationary state to another. The profile of dynamic pressure $\Pi$ follows this pattern, as well. The change of its form reminds of convergence to Dirac $\delta$ as $\alpha \to -1$. This is particularly convincing because the dynamic pressure does not exist in monatomic gases.

\begin{figure}[h]
    \centering
    \begin{subfigure}[b]{0.38\textwidth}
        \includegraphics[width=\textwidth]{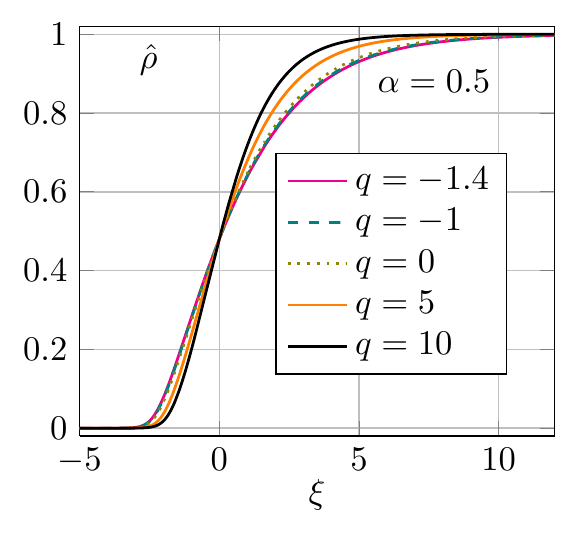}
        \label{fig:rho_param_2}
    \end{subfigure}
    ~ 
    \begin{subfigure}[b]{0.38\textwidth}
        \includegraphics[width=\textwidth]{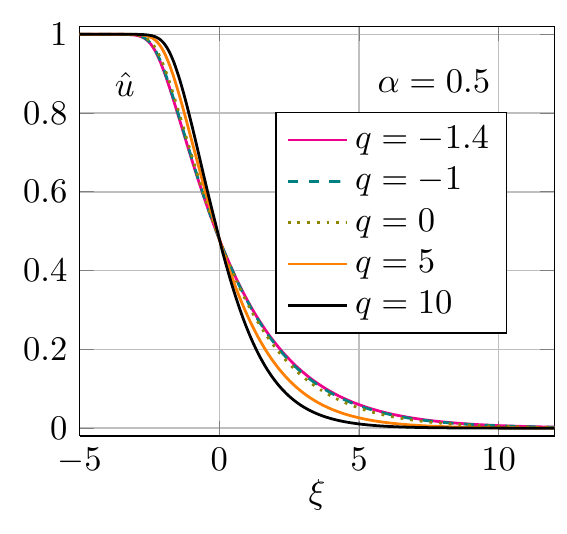}
        \label{fig:rho_param_2}
    \end{subfigure}
    ~ 
    \begin{subfigure}[b]{0.38\textwidth}
        \includegraphics[width=\textwidth]{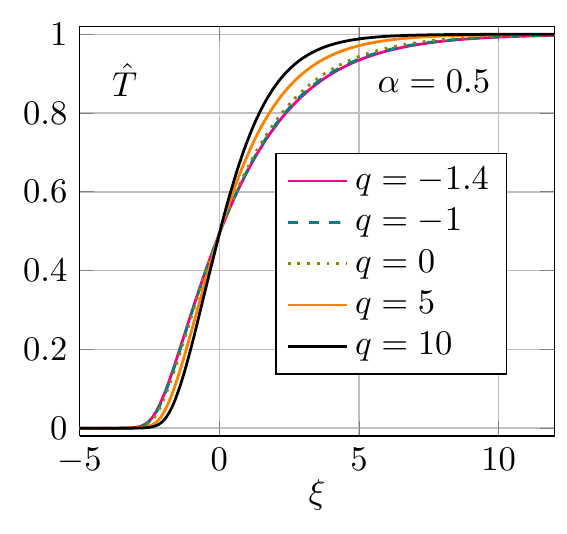}
        \label{fig:T_param_2}
    \end{subfigure}
    ~ 
    \begin{subfigure}[b]{0.38\textwidth}
        \includegraphics[width=\textwidth]{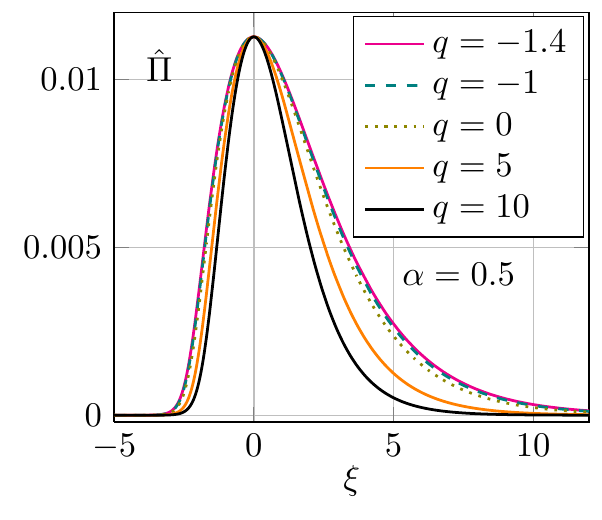}
        \label{fig:P_param_2}
    \end{subfigure}
    \caption{Shock profiles: $M_0=1.1$, $s = 0$, $\alpha = 0.5$, $\beta = 0$.}\label{fig:param_gen_1}
\end{figure}

Figure \ref{fig:param_gen_1} shows continuous shock profiles obtained using generalized source term \eqref{production term explicit general case-Pi}. Parameters of the model are chosen in a way that $\alpha$ and $\beta$ are fixed and satisfy the compatibility relation \eqref{ET_Compatibility}$_{1}$ with ET source term, while $s = 0$ is fixed. Parameter $q$ is varied and satisfy compatibility relation \eqref{ET_Compatibility}$_{2}$ with ET source term for $q = -1$. It is clear that the profiles change their shape with the increase of $q$ just as in the case of varying $s$---they become steeper. Moreover, the change is continuous and monotonous, so that ET profile ($q = -1$) lies between other profiles.

\subsection{Discontinuous shock structure}

Discontinuous shock structure, i.e. the shock structure with a sub-shock, appears when the shock speed exceeds the highest characteristic speed in upstream equilibrium, or equivalently:
\begin{equation}\label{MachCrit}
  M_{0} > M_{0}^{\ast} = \sqrt{\frac{5}{3} \frac{5 + 2 \alpha}{7 + 2 \alpha}}.
\end{equation}
In this case, the profile consists of a jump from upstream equilibrium \eqref{RH-Euler}$_{1}$ to intermediate state, followed by a continuous solution of shock structure equations \eqref{DLessShockEquations} that asymptotically tends to downstream equilibrium \eqref{RH-Euler}$_{2}$. Intermediate state is determined as a non-trivial solution of Rankine-Hugoniot equations for a complete system of governing equations \eqref{full system field eqs-Pi}, connecting $(\rho_{0}, u_{0}, T_{0}, \Pi_{0})$ with:
\begin{equation}\label{RH-Complete}
  \left(
    \begin{array}{c}
      \rho_{S} \\
      u_{S} \\
      T_{S} \\
      \Pi_{S} \\
    \end{array}
  \right)
  =
  \left(
    \begin{array}{c}
      \frac{4 M_{0}^{2} (7 + 2 \alpha)}{M_{0}^{2} (7 + 2 \alpha) + 5 (5 + 2 \alpha)} \\
      \frac{5 (5 + 2 \alpha)}{4 M_{0} (7 + 2 \alpha)} + \frac{M_{0}}{4} \\
      \frac{9 M_{0}^{4} (7 + 2 \alpha)^{2}
      + 2 M_{0}^{2} (5 + 2 \alpha) (7 + 2 \alpha) (37 + 16 \alpha) - 15 (5 + 2 \alpha)^{2}}{
      16 M_{0}^{2} (5 + 2 \alpha)^{2} (7 + 2 \alpha)} \\
      \frac{(1 + \alpha) \left(3 M_{0}^{4} (7 + 2 \alpha)^{2} - 2 M_{0}^{2} (5 + 2 \alpha) (7 + 2 \alpha) - 5 (5 + 2 \alpha)^{2} \right)}{2 (5 + 2 \alpha)^{2} \left(M_{0}^{2} (7 + 2 \alpha) +
      5 (5 + 2 \alpha) \right)} \\
    \end{array}
  \right).
\end{equation}
Inequality \eqref{MachCrit} determines the lower bound for upstream Mach number, whereas $M_{0}$ is not bounded from above. This fact will enrich our study with the analysis of shock profiles for different values of $M_{0}$.

\begin{figure}[h]
    \centering
    \begin{subfigure}[b]{0.38\textwidth}
        \includegraphics[width=\textwidth]{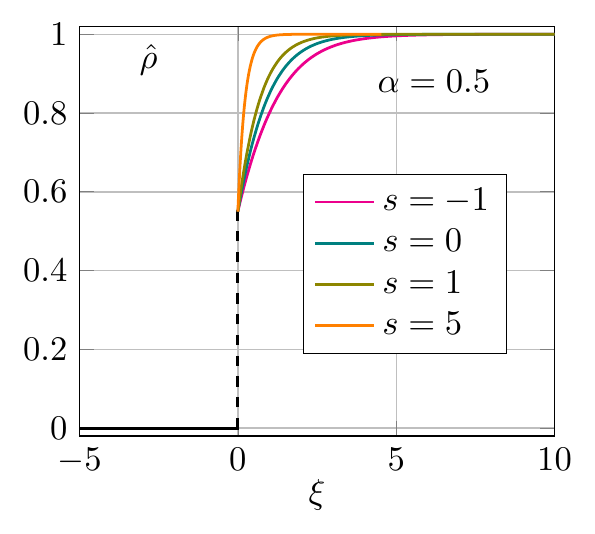}
        \label{fig:rho_sub_param_1}
    \end{subfigure}
    ~ 
    \begin{subfigure}[b]{0.38\textwidth}
        \includegraphics[width=\textwidth]{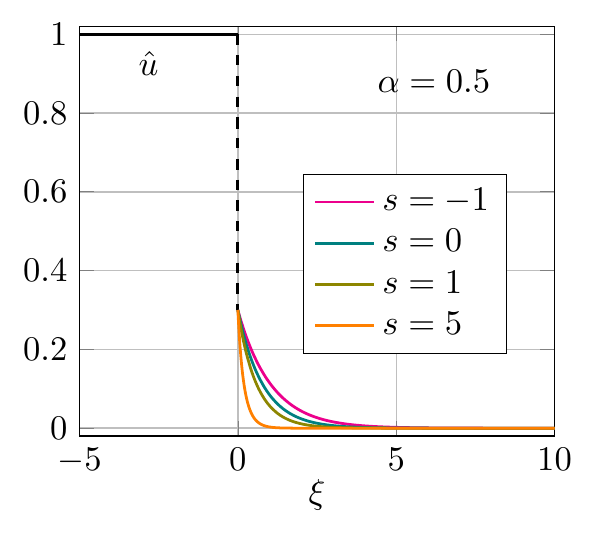}
        \label{fig:u_sub_param_1}
    \end{subfigure}
    ~ 
    \begin{subfigure}[b]{0.38\textwidth}
        \includegraphics[width=\textwidth]{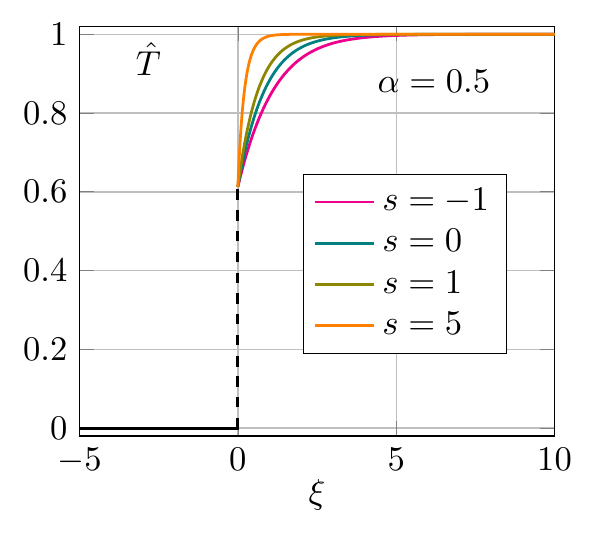}
        \label{fig:T_sub_param_1}
    \end{subfigure}
    ~ 
    \begin{subfigure}[b]{0.38\textwidth}
        \includegraphics[width=\textwidth]{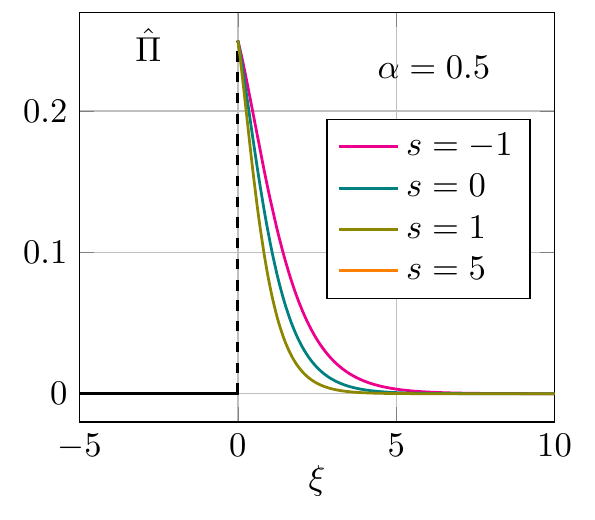}
        \label{fig:P_sub_param_1}
    \end{subfigure}
    \caption{Shock profiles: $M_0=1.5$, $\alpha = 0.5$}\label{fig:sub_param_1}
\end{figure}

Figure \ref{fig:sub_param_1} shows the dependence of shock profiles for different values of $s$, and fixed values of $\alpha = 0.5$ and $M_{0} = 1.5$. As in the continuous shock profiles, solutions become steeper with the increase of $s$. Note that sub-shocks are the same since $\alpha$ and $M_{0}$ are fixed.

\begin{figure}[h]
    \centering
    \begin{subfigure}[b]{0.38\textwidth}
        \includegraphics[width=\textwidth]{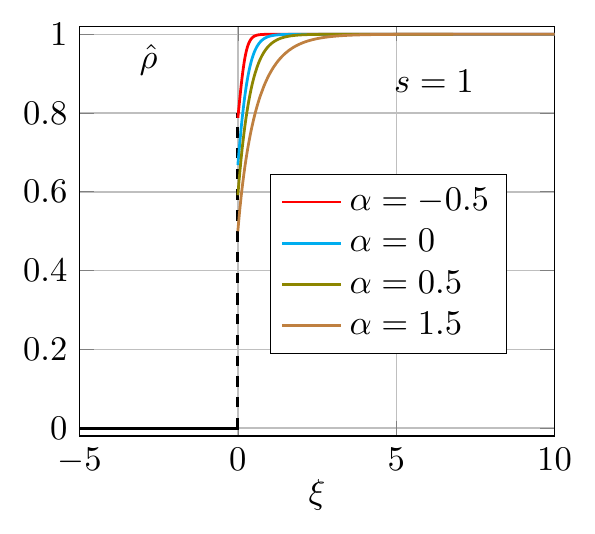}
        \label{fig:rho_sub_param_2}
    \end{subfigure}
    ~ 
    \begin{subfigure}[b]{0.38\textwidth}
        \includegraphics[width=\textwidth]{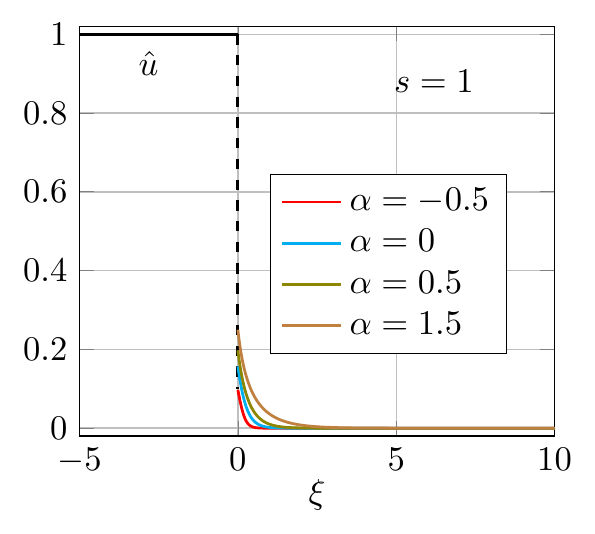}
        \label{fig:u_sub_param_2}
    \end{subfigure}
    ~ 
    \begin{subfigure}[b]{0.38\textwidth}
        \includegraphics[width=\textwidth]{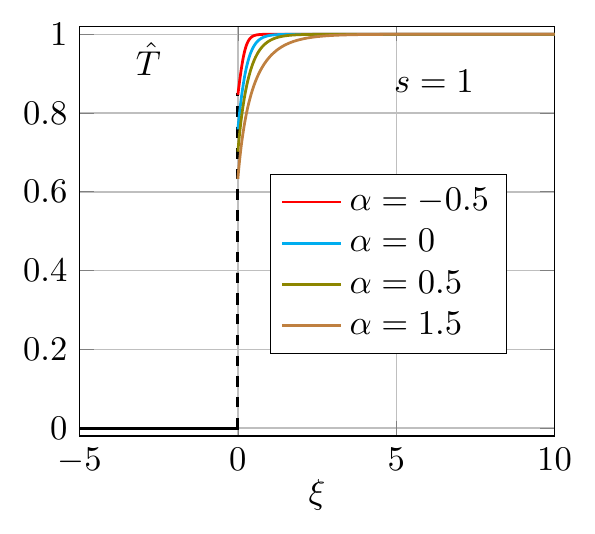}
        \label{fig:T_sub_param_2}
    \end{subfigure}
    ~ 
    \begin{subfigure}[b]{0.38\textwidth}
        \includegraphics[width=\textwidth]{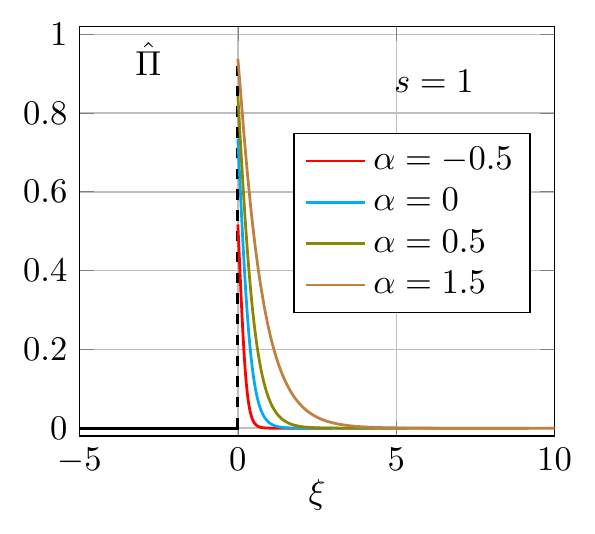}
        \label{fig:P_sub_param_2}
    \end{subfigure}
    \caption{Shock profiles: $M_0=2.0$, $s = 1$}\label{fig:sub_param_2}
\end{figure}

Figure \ref{fig:sub_param_2} presents the profiles for different values of $\alpha$, and fixed values for $s = 1.0$ and $M_{0} = 2.0$. Solutions for equilibrium state variables $\rho$, $u$ and $T$ are change in a different manner than solutions for $\Pi$. First, the jumps of equilibrium state variables are decreased for larger values of $\alpha$, whereas the jump of $\Pi$ is increased. On the other hand, profiles for smaller values of $\alpha$ are steeper than the profiles for larger values of $\alpha$ for all variables. One may draw a conclusion that when $\alpha \to -1$, corresponding to monatomic case, discontinuous shock structure of $\rho$, $u$ and $T$ tends to a pure shock, whereas the shock profile of $\Pi$ tends to vanish.

\begin{figure}[h]
    \centering
    \begin{subfigure}[b]{0.38\textwidth}
        \includegraphics[width=\textwidth]{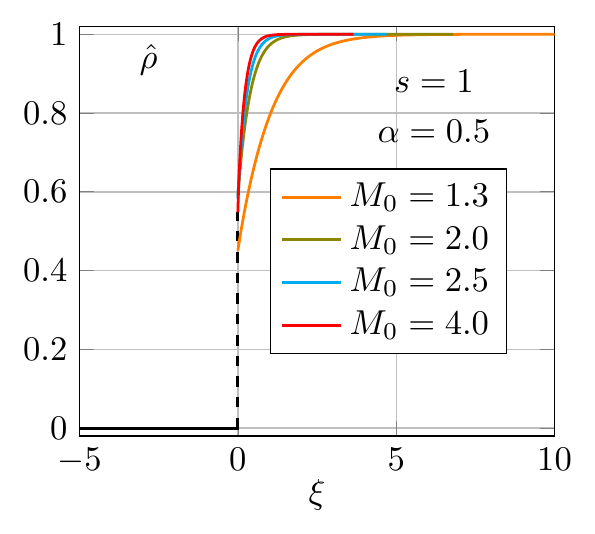}
        \label{fig:rho_sub_param_3}
    \end{subfigure}
    ~ 
    \begin{subfigure}[b]{0.38\textwidth}
        \includegraphics[width=\textwidth]{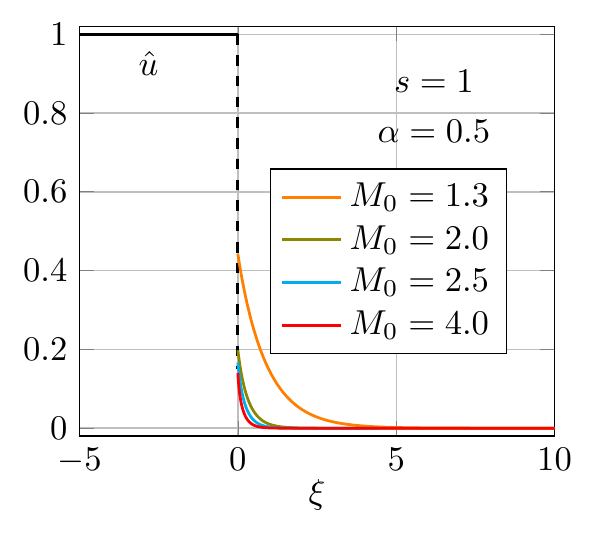}
        \label{fig:u_sub_param_3}
    \end{subfigure}
    ~ 
    \begin{subfigure}[b]{0.38\textwidth}
        \includegraphics[width=\textwidth]{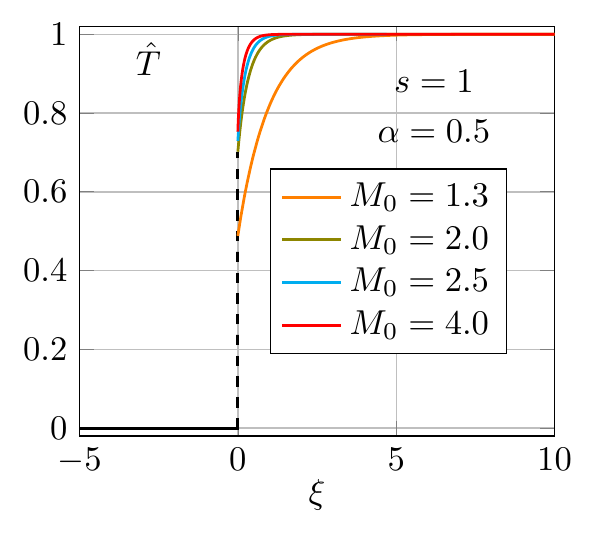}
        \label{fig:T_sub_param_3}
    \end{subfigure}
    ~ 
    \begin{subfigure}[b]{0.38\textwidth}
        \includegraphics[width=\textwidth]{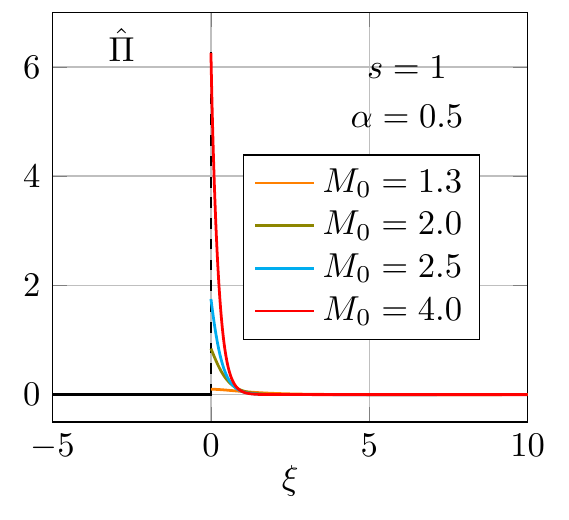}
        \label{fig:}
    \end{subfigure}
    \caption{Shock profiles: $\alpha = 0.5$, $s = 1$}\label{fig:sub_param_3}
\end{figure}

Figure \ref{fig:sub_param_3} shows the profiles for different values of $M_{0}$, and fixed values of $\alpha = 0.5$ and $s = 1.0$. Continuous parts of the shock profiles of all state variables become steeper with the increase of Mach number.

\begin{figure}[h]
    \centering
    \begin{subfigure}[b]{0.38\textwidth}
        \includegraphics[width=\textwidth]{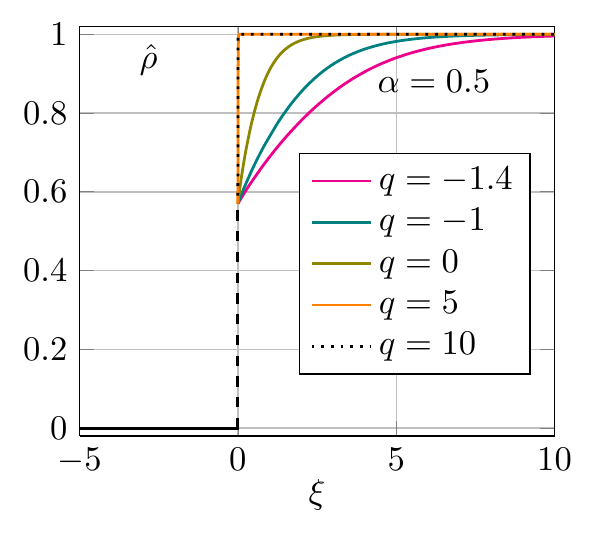}
        \label{fig:rho_sub_param_3}
    \end{subfigure}
    ~ 
    \begin{subfigure}[b]{0.38\textwidth}
        \includegraphics[width=\textwidth]{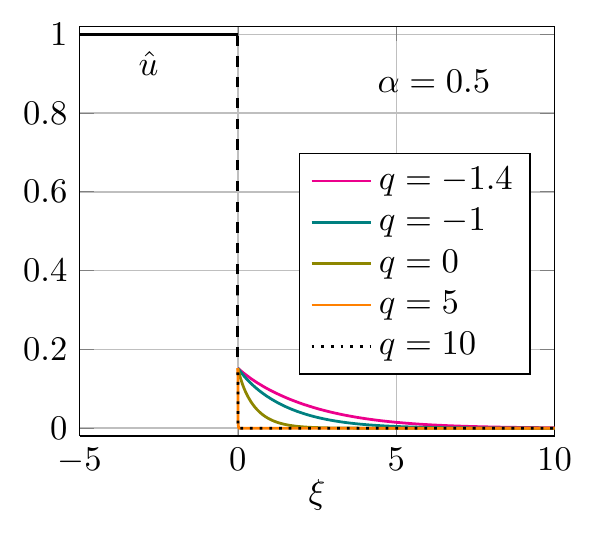}
        \label{fig:u_sub_param_3}
    \end{subfigure}
    ~ 
    \begin{subfigure}[b]{0.38\textwidth}
        \includegraphics[width=\textwidth]{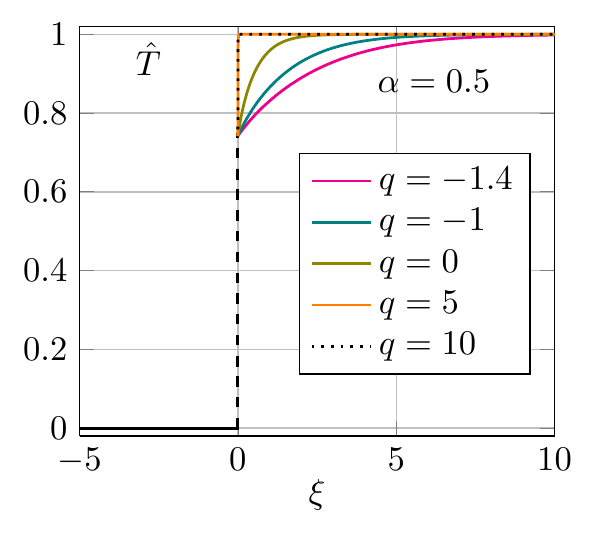}
        \label{fig:T_sub_param_3}
    \end{subfigure}
    ~ 
    \begin{subfigure}[b]{0.38\textwidth}
        \includegraphics[width=\textwidth]{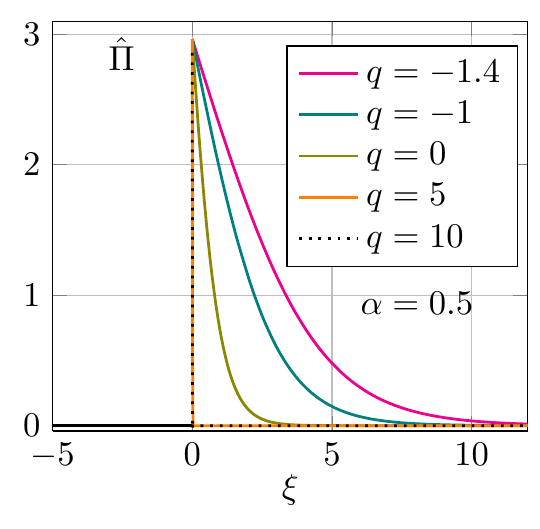}
        \label{fig:}
    \end{subfigure}
    \caption{Shock profiles: $M_{0} = 3.0$, $s = 0$, $\alpha = 0.5$, $\beta = 0$.}
    \label{fig:sub_param_gen_1}
\end{figure}

Figure \ref{fig:sub_param_gen_1} shows discontinuous shock profiles obtained for using generalized source term \eqref{production term explicit general case-Pi}. Parameters are chosen in the same way as in the case of continuous profiles. One may observe again that ET profile ($q = -1$) is embedded within other profiles. Furthermore, with the increase of $q$ discontinuous profiles converge to a ``pure'' jump connecting upstream and downstream equilibria, while jumps (``up'' and ``down'') of dynamic pressure tend to coincide.

\section{Conclusion}

In this paper we analyzed polyatomic gases starting from the kinetic viewpoint, through the application of maximum entropy principle. Attention was restricted to the model with 6 macroscopic fields---the simplest physically reasonable model that captures non-equilibrium phenomena. Moreover, it is one of the rare cases in which MEP admits non-linear closure. Internal degrees of freedom of the molecules were described using single continuous real variable.

Within this framework, we first determined the velocity distribution function as a solution of the variational problem with constraints---the problem of entropy maximization. After that, through the analysis of non-equilibrium part of the entropy, the dynamic pressure was recognized as a non-equilibrium part of the trace of the pressure tensor. This led to a macroscopic description of polyatomic gases and derivation of field equations in the form of hyperbolic systems of balance laws. Principal novelty in this part of the study consists in computation of the source term using the complete collision operator, rather than BGK model. The non-linear source term contains parameters that make it akin to adapt to real physical situations. However, this analysis raised an important question about restrictions on the collision cross section in the kinetic theory. It was revealed that violation of Galilean invariance at molecular level does not imply the violation of the same condition at macroscopic level---the non-invariance is wiped-out through integration over velocity space. Actually, it was realized that this is the only way by which we can match our results with the source term obtained through non-linear closure within extended thermodynamics. Final part of the paper was devoted to the parametric study of the shock structure. It revealed certain interesting properties related to the structure of molecules (degrees of freedom) and the parameters in the source term. As indicated above, their presence provide a whole spectrum of different shock profiles, and their values can be chosen such that results of simulations match the real data obtained by some other mean.

This study of polyatomic gases with dynamic pressure provides a contribution to some recent discussions within similar framework. However, it also opens some new problems and possible fields of study. For example, our study of the non-linear source term is just a scratch that reveals complex relations between descriptions of the same phenomena at different scales and different levels of accuracy. This question deserves much deeper analysis, especially due to close relations between kinetic theory gases and extended thermodynamics. First of all, the question of Galilean invariance appeared to be rather delicate and should be treated with particular care in further studies. Furthermore, in the structure of source term there appears a singularity that prevents simple asymptotic treatment of the monatomic case. This problem does not appear in extended thermodynamics, where phenomenological coefficient hides deeper structural information. Question of passage from polyatomic to monatomic case in the kinetic theory is more subtle than in extended thermodynamics. It cannot be resolved by the sole analysis of macroscopic equations and should be addressed in order to complete the picture of relation between macro and meso scale.

\appendix
\section{Calculation of the source term}

In this Appendix, we detail the calculation of the production term \eqref{production term} for the specific choice of the distribution function $f$, the weight function $\vfi$ and the cross section $\mathcal{B}$. Namely, for the distribution function we take the one that comes out as a solution of the MEP problem i.e. \eqref{solution MEP}, the weight is chosen to be $\vfi=I^\alpha$, $\alpha>-1$, and the cross section is
\beqs
\mathcal{B}(\bv, \bv_*, I, I_*, r, R, \bom) = K \,R^s \left| \bv-\bv_* \right|^{2s},
\eeqs
where $K$ is an appropriate dimensional constant and  the parameter $s$ satisfies the overall assumption $s>-\frac{3}{2}$.

As usual, computations are facilitated when one deals with the peculiar velocity
\beq\label{peculiar velocity}
\bc=\bv-\bu.
\eeq
Indeed, let us rewrite the collision transformations from  Section \ref{section: coll transf}. As one expects, they take the same form as for the velocity $\bv$, since in general the problem should not depend upon the macroscopic velocity $\bu$. The conservation laws  at a microscopic level \eqref{microscopic CL} in terms of $\bc$ read:
\beq\label{micropcopic CL for peculiar velocity}
\begin{split}
m \, \bc' + m \, \bc'_* &= m \, \bc + m \, \bc_* ,\\
\tfrac{m}{2} \left| \bc' \right|^2 + \tfrac{ m}{2} \left| \bc'_* \right|^2 + I' + I'_* &= \tfrac{m}{2} \left| \bc \right|^2 + \tfrac{m}{2}  \left| \bc_* \right|^2 + I + I_*.
\end{split}
\eeq
They allow to write the pre-collisional velocities with the help of the parameter $\bom \in S^2$ as follows
\beq\label{collisional rules for peculiar velocity}
\bc'=  \frac{\bc+\bc_*}{2} + \koren \, \Tomc,\qquad
\bc'_* =  \frac{\bc+\bc_*}{2} - \koren \, \Tomc,
\eeq
where
\beq\label{E preko c}
E= \tfrac{m}{4}   \left| \bc'-\bc'_* \right|^2  + I' + I'_*= \tfrac{m}{4} \left| \bc-\bc_* \right|^2  + I + I_*,
\eeq
and $T_{\bom} \left[  \mathbf{y} \right] = \mathbf{y} - 2 \lp \bom \cdot \mathbf{y} \rp \bom$ (for any $\mathbf{y} \in \bR$).\\

Let us now perform required calculation:
\beqs
\sum_{i=1}^3 {\Pi}_{ii} = \int_{\bR \times \bRp} m \left| \bv \right|^2 Q(\hat{f}, \hat{f})(\bv, I) I^{\alpha} \, \md I \md \bv.
\eeqs

First, note that by the conservative properties of the weak form \eqref{anuliranje weak form}$_{1,2}$, the production term reduces to
\begin{multline*}
\sum_{i=1}^3 {\Pi}_{ii} = \int_{\bR \times \bRp} m \left| \bc \right|^2 Q(\hat{f}, \hat{f})(\bv, I) I^{\alpha} \, \md I \md \bv \\ =
m \,  K  \int_{\bR \times \bRp \times \bR \times \bRp \times [0,1]^2 \times S^2 } \left|\bc\right|^2  \lp \hat{f}' \hat{f}'_* - \hat{f} \hat{f}_*   \rp \left| \bv-\bv_* \right|^{2s} (1-R) R^{s+\frac{1}{2}} \, \md \bom \, \dsvi.
\end{multline*}
After performing the change of variables $ \lp \bv, \bv_*, I, I_*, r, R, \bom \rp \mapsto  \lp \bv', \bv'_*, I', I'_*, r', R', \bom \rp$ with the Jacobian \eqref{jacobian non prime -> prime} in the first  integral and passing to the relative velocities by means of the change of variables $\lp \bv, \bv_* \rp \mapsto \lp \bc, \bc_* \rp$ with unit Jacobian, the production term takes the form:
\begin{multline*}
\sum_{i=1}^3 {\Pi}_{ii} = m \, \cnst^2 \, K  \int_{\bR \times \bRp \times \bR \times \bRp \times [0,1]^2 \times S^2 } \lp \left| \bc' \right|^2 - \left|\bc\right|^2 \rp
 e^{-\frac{3\rho}{2 \sum_{i=1}^3 p_{ii}} \lp \left|  \bc \right|^2 + \left|  \bc_* \right|^2 \rp - \frac{\alpha+1}{m \, \lp e - \frac{1}{2 \rho} \sum_{i=1}^3 p_{ii} \rp} \lp I + I_* \rp } \\ \times \left| \bc-\bc_* \right|^{2s} (1-R) R^{s+\frac{1}{2}} \, \md \bom \, \dsvic,
\end{multline*}
where  $\cnst$   is given by
\beqs
\cnst =  \frac{\rho}{m} \lp \frac{3 \rho}{2 \pi \sum_{i=1}^3 p_{ii}}  \rp^{\frac{3}{2}} \lp \frac{\alpha+1}{m \, \lp e - \frac{1}{2 \rho} \sum_{i=1}^3 p_{ii} \rp} \rp^{\alpha+1} \frac{1}{\Gamma\left[ \alpha+1\right]}.
\eeqs
It is convenient to perform the following change of variables with unit Jacobian
\begin{equation}\label{c,cs --> g, G}
\lp \bc, \bc_* \rp \mapsto \lp \bg=\bc-\bc_*, \bG= \frac{\bc+\bc_*}{2} \rp,
\end{equation}
which allows to express
\beqs
\bc = \bG + \frac{1}{2} \bg, \qquad
\bc_* = \bG - \frac{1}{2} \bg.
\eeqs
By microscopic conservation law of momentum \eqref{micropcopic CL for peculiar velocity}$_1$, it holds $\frac{\bc'+\bc'_*}{2}=:\bG'=\bG$. Also, using \eqref{collisional rules for peculiar velocity}, we can see that
\beq\label{polyA: g' preko c,cs}
\bg':= \bc'-\bc'_* = 2 \koren \Tomc = 2 \koren \left[ \frac{\bg}{\left| \bg \right|} \right],
\eeq
and therefore
\beqs
\bc' = \bG + \frac{1}{2} \bg', \qquad
\bc'_* = \bG - \frac{1}{2} \bg'.
\eeqs
We are led to rewrite the terms under the integration sign using new variables
\begin{align*}
\left| \bc' \right|^2 - \left| \bc \right|^2 &=  \bG \cdot \left( \bg' - \bg \right) + \frac{1}{4} \lp R - 1 \rp \left| \bg \right|^2 + \frac{R}{m} \lp I+I_* \rp.
\end{align*}
The terms appearing in the power of the exponential read:
\begin{align}
\left| \bc \right|^2 + \left| \bc_* \right|^2 = \left| \bG + \frac{1}{2} \bg \right|^2 + \left| \bG - \frac{1}{2} \bg \right|^2
&= 2 \left| \bG \right|^2 + \frac{1}{2} \left| \bg \right|^2. \label{c^2+ cs^2 in terms of g, G}
\end{align}
Passage to velocities $\bg$ and $\bG$ and insertion of the formulas above yield
\begin{multline*}
\sum_{i=1}^3 {\Pi}_{ii} = m \, \cnst^2 \, K  \int_{\bR \times \bRp \times \bR \times \bRp \times [0,1]^2 \times S^2 } \lp  \bG \cdot \left( \bg' - \bg \right) + \frac{1}{4} \lp R - 1 \rp \left| \bg \right|^2 + \frac{R}{m} \lp I+I_* \rp \rp \\ \times
 e^{-\frac{3\rho}{2 \sum_{i=1}^3 p_{ii}} \lp 2 \left|  \bG \right|^2 + \frac{1}{2} \left|  \bg \right|^2 \rp - \frac{\alpha+1}{m \, \lp e - \frac{1}{2 \rho} \sum_{i=1}^3 p_{ii} \rp} \lp I + I_* \rp } \left| \bg \right|^{2s} (1-R) R^{s+\frac{1}{2}} \, \md \bom \, \dsvig.
\end{multline*}
Integration with respect to $\bG$ leads to:
\begin{multline*}
\sum_{i=1}^3 {\Pi}_{ii} = m \, \cnst^2 \, K \, \lp \frac{3 \rho}{\sum_{i=1}^3 p_{ii}} \rp^{-3/2} \pi^{3/2} \int_{\bR \times \bRp  \times \bRp \times [0,1]^2 \times S^2 } \lp   \frac{1}{4} \lp R - 1 \rp \left| \bg \right|^2 + \frac{R}{m} \lp I+I_* \rp \rp \\ \times
 e^{-\frac{3 \rho}{4 \sum_{i=1}^3 p_{ii}}  \left|  \bg \right|^2  - \frac{\alpha+1}{m \, \lp e - \frac{1}{2 \rho} \sum_{i=1}^3 p_{ii} \rp} \lp I + I_* \rp } \left| \bg \right|^{2s} (1-R) R^{s+\frac{1}{2}} \, \md \bom \, \mathrm{d}r \, \mathrm{d}R \, \mathrm{d}I_* \,  \mathrm{d}I \,  \mathrm{d}\bg.
\end{multline*}
Straightforward integration with respect to variables $\bom$ and $r$ yields
\begin{multline*}
\sum_{i=1}^3 {\Pi}_{ii} = m \, \cnst^2 \, K \, \lp \frac{3 \rho}{\sum_{i=1}^3 p_{ii}} \rp^{-3/2} 4 \pi^{5/2} \int_{\bR \times \bRp  \times \bRp \times [0,1] } \lp   \frac{1}{4} \lp R - 1 \rp \left| \bg \right|^2 + \frac{R}{m} \lp I+I_* \rp \rp \\ \times
 e^{-\frac{3\rho}{4 \sum_{i=1}^3 p_{ii}}  \left|  \bg \right|^2  - \frac{\alpha+1}{m \, \lp e - \frac{1}{2 \rho} \sum_{i=1}^3 p_{ii} \rp} \lp I + I_* \rp } \left| \bg \right|^{2s} (1-R) R^{s+\frac{1}{2}} \, \mathrm{d}R \, \mathrm{d}I_* \,  \mathrm{d}I \,  \mathrm{d}\bg.
\end{multline*}
Next, we integrate with respect to $R$ and we obtain
\begin{multline*}
\sum_{i=1}^3 {\Pi}_{ii} = m \, \cnst^2 \, K \, \lp \frac{3 \rho}{\sum_{i=1}^3 p_{ii}} \rp^{-3/2}  \pi^{5/2} \frac{16}{ \lp 2s+5 \rp \lp 2s+7 \rp} \\ \times \int_{\bR \times \bRp  \times \bRp } \lp  - \frac{1}{\lp 2s+3 \rp} \left| \bg \right|^2 + \frac{1}{m} \lp I+I_* \rp \rp
 e^{-\frac{3 \rho}{4 \sum_{i=1}^3 p_{ii}}  \left|  \bg \right|^2  - \frac{\alpha+1}{m \, \lp e - \frac{1}{2 \rho} \sum_{i=1}^3 p_{ii} \rp} \lp I + I_* \rp } \left| \bg \right|^{2s}  \, \mathrm{d}I_* \,  \mathrm{d}I \,  \mathrm{d}\bg.
\end{multline*}
Furthermore, we proceed the integration with respect to $I$ and $I_*$
\begin{multline*}
\sum_{i=1}^3 {\Pi}_{ii} = m \, \cnst^2 \, K \, \lp \frac{3 \rho}{\sum_{i=1}^3 p_{ii}} \rp^{-3/2}  \pi^{5/2} \frac{16}{ \lp 2s+5 \rp \lp 2s+7 \rp} \lp \frac{m \, \lp e - \frac{1}{2 \rho} \sum_{i=1}^3 p_{ii} \rp }{\alpha+1 } \rp^2 \\ \times \int_{\bR} \lp  - \frac{1}{\lp 2s+3 \rp}  \left| \bg \right|^2 +  2 \frac{\lp e - \frac{1}{2 \rho} \sum_{i=1}^3 p_{ii} \rp}{\lp \alpha+1 \rp} \rp
 e^{-\frac{3 \rho}{4 \sum_{i=1}^3 p_{ii}}  \left|  \bg \right|^2 } \left| \bg \right|^{2s}  \,  \mathrm{d}\bg.
\end{multline*}
Finally, after integration with respect to $\bg$ we enclose the calculation of the production term
\begin{multline*}
\sum_{i=1}^3 {\Pi}_{ii} = K \, \frac{\rho^2}{m}\, \pi^{1/2} \, \frac{2^{2s+6} \, \Gamma\left[ s + \frac{3}{2} \right] }{\lp 2 s+ {5} \rp \lp 2 s+ {7} \rp}  \lp \frac{1} {3\rho} \sum_{i=1}^3 p_{ii} \rp^{s} \frac{1}{\Gamma\left[ \alpha + 1 \right]^2} \lp  \frac{\alpha+1 }{m \, \lp e - \frac{1}{2 \rho} \sum_{i=1}^3 p_{ii} \rp } \rp^{2 \alpha} \\ \times \lp - \frac{1} {3\rho} \sum_{i=1}^3 p_{ii} + \frac{1}{\alpha+1} \lp e - \frac{1}{2 \rho} \sum_{i=1}^3 p_{ii} \rp \rp.
\end{multline*}

\end{document}